\documentclass{aastex}
\usepackage{spr-astr-addons}             
\usepackage{lscape, morefloats, graphicx, float}
   
\RequirePackage{color}                   

\begin{document}

\title{Metallicity Calibration and Photometric Parallax Estimation: II. SDSS photometry}
\slugcomment{Not to appear in Nonlearned J., 45.}
\shorttitle{Metallicity Calibration and Photometric Parallax Estimation}
\shortauthors{S. Tun{\c c}el G\"u{\c c}tekin, Bilir, S., Karaali, S., O. Plevne, Ak, S., T. Ak, Z. F. Bostanc\i}

\author{S. Tun{\c c}el G\"u{\c c}tekin \altaffilmark{1}}
\altaffiltext{1}{Istanbul University, Graduate School of Science and Engineering, 
Department of Astronomy and Space Sciences, 34116, Beyaz\i t, Istanbul, Turkey\\
\email{sabihatuncel@gmail.com}}

\author{S. Bilir \altaffilmark{2}} 
\altaffiltext{2}{Istanbul University, Faculty of Science, Department 
of Astronomy and Space Sciences, 34119 University, Istanbul, Turkey\\}
\and
\author{S. Karaali \altaffilmark{2}} 
\altaffiltext{2}{Istanbul University, Faculty of Science, Department 
of Astronomy and Space Sciences, 34119 University, Istanbul, Turkey\\}
\and
\author{O. Plevne \altaffilmark{1}}
\altaffiltext{1}{Istanbul University, Graduate School of Science and Engineering, 
Department of Astronomy and Space Sciences, 34116, Beyaz\i t, Istanbul, Turkey\\}
\and
\author{S. Ak \altaffilmark{2}}
\altaffiltext{2}{Istanbul University, Faculty of Science, Department 
of Astronomy and Space Sciences, 34119 University, Istanbul, Turkey\\}
\and
\author{T. Ak\altaffilmark{2}} 
\altaffiltext{2}{Istanbul University, Faculty of Science, Department 
of Astronomy and Space Sciences, 34119 University, Istanbul, Turkey\\}
\and
\author{Z. F. Bostanc\i\altaffilmark{2}} 
\altaffiltext{2}{Istanbul University, Faculty of Science, Department 
of Astronomy and Space Sciences, 34119 University, Istanbul, Turkey\\}

\begin{abstract} 
We used the updated [Fe/H] abundances of 168 F-G type dwarfs and calibrated them to a third order polynomial in terms of reduced ultraviolet excess, $\delta_{0.41}$ defined with $ugr$ data in the SDSS. We estimated the $M_g$ absolute magnitudes for the same stars via the re-reduced {\em Hipparcos} parallaxes  and calibrated the absolute magnitude offsets, $\Delta M_g$, relative to the intrinsic sequence of Hyades to a third order polynomial in terms of $\delta_{0.41}$. The ranges of the calibrations are $-2<$[Fe/H]$\leq$0.3 dex and $4<M_g\leq6$ mag. The mean of the residuals and the corresponding standard deviation for the metallicity calibration are 0 and 0.137 mag; while, for the absolute magnitude calibration they are 0 and 0.179 mag, respectively. We applied our procedures to 23,414 dwarf stars in the Galactic field with the Galactic coordinates $85^{\circ}\leq b\leq90^{\circ}$, $0^{\circ}\leq l\leq360^{\circ}$ and size 78 deg$^{2}$. We estimated absolute magnitude $M_g$ dependent vertical metallicity gradients as a function of vertical distance $Z$. The gradients are deep in the range of $0<Z\leq5$ kpc, while they are very small positive numbers beyond $Z=5$ kpc. All dwarfs with $5<M_g\leq6$ mag are thin-disc stars and their distribution shows a mode at $(g-r)_0\approx 0.38$ mag, while the absolute magnitudes $4<M_g\leq5$ are dominated by thick disc and halo stars, i.e. the apparently bright ones ($g_0\leq18$ mag) are thick-disc stars with a mode at $(g-r)_0\sim0.38$ mag, while the halo population is significant in the faint stars ($g_0>18$ mag). 
\end{abstract}

\keywords{Galaxy: disc, Galaxy: halo, stars: abundances, stars: distances}    

\section{Introduction}
Metallicity distribution of  stars in our Galaxy has been used as an indicator for understanding its formation. The pioneer work related to the formation of our Milky Way Galaxy is the one of \citet*[][hereafter ELS]{ELS62} who argued that the Galaxy collapsed in a free-fall time, i.e. 2$\times$10$^{8}$ Gyr. However, this argument was in contradiction with the metallicity gradients observed at least in some of the components of the Galaxy,  which indicate a dissipative collapse in a much longer time-scale than the predicted one of ELS. Many studies confirmed this new idea \citep*[cf.][and references therein]{Karaali04, Bilir06, Bilir08a}. Additionally, we know that at least some of the components of the Galaxy were formed from merger or accretion of numerous fragments, such as dwarf-type galaxies \citep[cf.][]{Searle78, Freeman02}. A positive iron abundance gradient in the increasing vertical direction of the Galactic halo is a typical feature for the mentioned mergers and accretions. Despite these studies, the researchers still investigate the formation -and evolution- of the Galaxy by using the recent data, which are more accurate. 

Metallicity of a star can be determined spectroscopically or photometrically. The first procedure requires high-resolution spectra, hence it can be applied only to nearby stars. While the second one is valid for distant stars as well. The invent of large telescopes which provide high resolution increased the accuracy of metallicity estimation. 
 
The distance to a star is important in the kinematical investigation of the Galaxy. It can be estimated by using trigonometric or photometric parallaxes. The estimation of the trigonometric parallax of a star is limited with distance, i.e. it can be applied only to nearby stars. While, photometric parallax can be applied to distant stars as well. The main source for the trigonometric parallax is the {\it Hipparcos} data \citep{ESA97, vanLeeuwen07}. Photometric parallax of a star can be provided by combination of its apparent and absolute magnitudes \citep[cf.][]{Bilir08b, Bilir09}. However, estimation of the absolute magnitude of a star is another problem. The procedure  usually used for its estimation is based on its offset from a standard sequence, such as the absolute magnitude-colour diagram of the Hyades cluster. Examples can be found in \citet{Laird88, Nissen91, Karaali03a, Karaali05, Karatas06}. The procedure of \citet{Ivezic08} is different than the cited ones, i.e. they calibrated the [Fe/H] to a third order polynomial in terms of $(g-r)_0$, $(u-g)_{0}$ and their combination. Also, they calibrated the absolute magnitude $M_r$ instead of $M_g$. Additionally, the standard sequence used in their calibration consists of a theoretical third order polynomial of the colour $(g-i)_0$. Alternative methods are given in some studies in the literature \citep{Phleps00, Chen01, Siegel02}.
  
In our previous study \citep[][hereafter Paper I]{Guctekin16}, we used the $UBV$ data and calibrated the iron abundance [Fe/H] and absolute magnitude offset $\Delta M_V$ in terms of  the reduced ultraviolet (UV) excess, $\delta_{0.6}(U-B)$. In this study, our calibrations will be based on the $ugr$ data of Sloan Digital Sky Survey \citep[SDSS,][]{York00} which has a wider application area, i.e. [Fe/H] and $\Delta M_g$ will be calibrated in terms of ultra-violet excess in the $ugr$ system, where $M_g$ is the absolute magnitude of a star corresponding to the $g$-band. However, here we have a difficulty, i.e. the spaces occupied by stars observed with {\em Hipparcos} \citep{ESA97} and SDSS are different. The solution of this problem is to transform the $UBV$ data of the sample stars in Paper I to the $ugr$ data and adopt the same procedure for our purpose. Thus, we should extend our calibrations to stars observed with SDSS at larger distances. Also, the application of our new calibrations will be carried out in this study. We organized the paper as follows. The data are presented in Section 2. The procedure and its application are given in Section 3 and Section 4, respectively. Finally, Section 5 is devoted to Summary and Discussion.

\section{Data}
The data consist of $g_0$ magnitudes, $(g-r)_0$ and $(u-g)_0$ colours, [Fe/H] iron abundances, $M_g$ absolute magnitudes of a sample of 168 stars; and the $(u-g)_0\times(g-r)_0$ two-colour diagram and $M_g\times(g-r)_0$ sequence of the Hyades cluster, which are provided as explained in the following. The subscript ``0'' denotes the de-reddened magnitudes and colours. The sample is the same as in Paper I, i.e. they are F and G type dwarfs with $5310<T_{eff}(K)<7300$ and $\log g>4$ (cgs). Hence, their iron abundances are identical with the ones used in Paper I.

We used the transformation equations of \citet{Chonis08} (Eqs. 1-5) and obtained their inverse ones (Eqs. 6-8) as in the following:

\begin{equation}
B_0=g_0+0.327(g-r)_0+0.216.\\
\end{equation}

\begin{equation}
V_0=g_0-0.587(g-r)_0-0.011.\\
\end{equation}

\begin{equation}
R_0=r_0-0.272(r-i)_0-0.159.\\
\end{equation}

\begin{equation}
I_0=i_0-0.337(r-i)_0-0.370.\\
\end{equation}

\begin{equation}
U_0=u_0-0.854.\\
\end{equation}

\begin{equation}
g_0=V_0+0.642(B-V).\\
\end{equation}

\begin{equation}
(u-g)_0=(U-B)_0+0.358(B-V)_0+0.989.\\
\end{equation}

\begin{equation}
(g-r)_0=1.094(B-V)_0-0.248.\\
\end{equation}

The $g_0$, $(g-r)_0$ and $(u-g)_0$ data are given in Table 1. The range of the $(g-r)_0$ colour index is $0.15<(g-r)_0<0.50$ mag. The iron abundances of the sample stars which are taken from Paper I \citep[originally from][]{Bensby14, Nissen10, Reddy06, Venn04} are also given in Table 1 and their distribution is presented in Fig. 1. The trigonometric parallaxes are provided from the re-reduced {\em Hipparcos} catalogue \citep{vanLeeuwen07}; 84\% of their relative parallax errors are less than 0.1 (Fig. 2). The absolute magnitudes are evaluated by the following Pogson equation:

\begin{equation}
g_0-M_g=5\times \log d-5.
\end{equation}
where $d$ is the distance to the star estimated via its trigonometric parallax. The distribution of absolute magnitudes are shown in Fig. 3. The range of the absolute magnitudes is $3.5\leq M_g\leq6.5$ mag. 

\begin{figure}[t]
\begin{center}
\includegraphics[scale=0.85, angle=0]{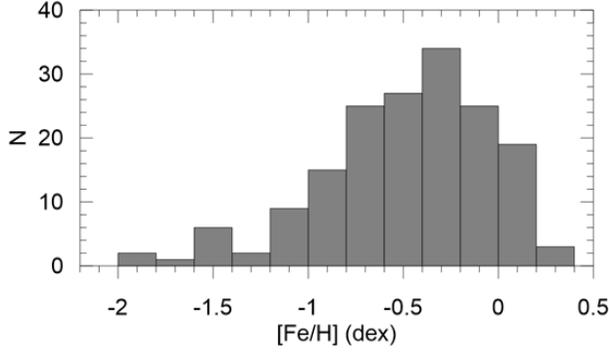}
\caption[] {Metallicity distribution for 168 sample stars.}
\end{center}
\end{figure}

\begin{figure}[t]
\begin{center}
\includegraphics[scale=0.85, angle=0]{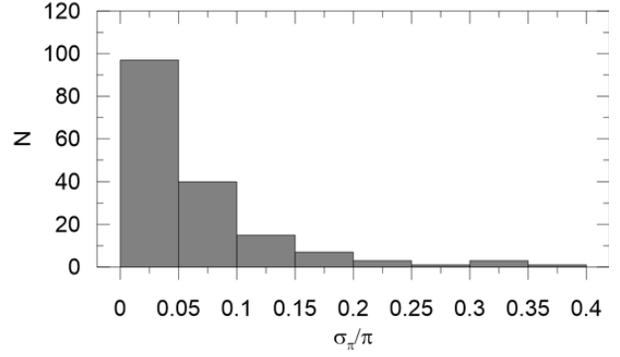}
\caption[] {Distribution of the relative parallax errors for 168 sample stars.} 
\end{center}
\end {figure}

\begin{figure}[H]
\begin{center}
\includegraphics[scale=0.85, angle=0]{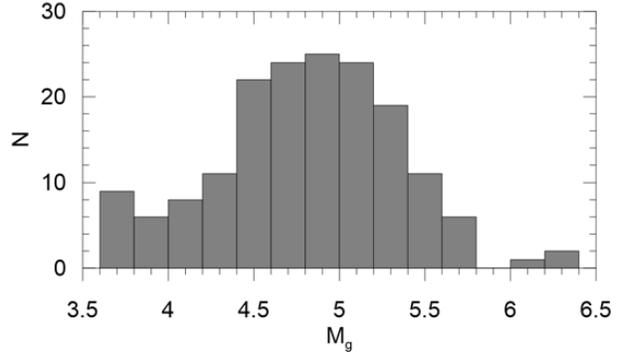}
\caption[] {Original absolute magnitude distribution for 168 sample stars.}
\end{center}
\end{figure}

\begin{table*}
\setlength{\tabcolsep}{1.5pt} 
{\scriptsize
\caption{Atmospheric and photometric parameters for 168 sample stars used in the calibrations. The columns give: ID, Hipparcos number, equatorial coordinates ($\alpha, \delta$), trigonometric parallax ($\pi$), effective temperature ($T_{eff}$), surface gravity ($\log g$), iron abundance [Fe/H], apparent magnitude ($V_0$), colour indices ($(U-B)_0,(B-V)_0$), apparent magnitude ($g_0$), and colour indices ($(u-g)_0,(g-r)_0$).} 
\begin{center}
\begin{tabular}{cccccccccccccc}
\hline
ID & Hip & $\alpha_{2000}$ &$\delta_{2000}$ & $\pi$  & $T_{eff}$ & $\log g$ & [Fe/H] & $V_o$ & $(U-B)_o$ & $(B-V)_o$ & $g_o$ &  $(u-g)_o$ & $(g-r)_o$\\
   &     & (hh:mm:ss.ss) & (dd:mm:ss.s) & (mas)  & (K) & (cgs)& (dex) & (mag) & (mag) & (mag) & (mag) & (mag) & (mag)\\
\hline
1 & 493    & 00 05 54.70 & +18 14 06.0 & 26.93 & 5826 & 4.36 & -0.29 & 7.435 &  0.036 & 0.555 & 7.656 &  1.224 & 0.359 \\
2 & 910    & 00 11 15.86 & -15 28 04.7 & 53.34 & 6289 & 4.17 & -0.33 & 4.886 & -0.016 & 0.489 & 5.065 &  1.148 & 0.287 \\
3 & 1599   & 00 20 04.26 & -64 52 29.3 &116.46 & 5932 & 4.33 & -0.19 & 4.225 &  0.017 & 0.571 & 4.457 &  1.210 & 0.377 \\
4 & 1976   & 00 25 01.42 & -30 41 51.7 & 21.27 & 5982 & 4.32 & +0.19 & 7.548 &  0.152 & 0.611 & 7.805 &  1.360 & 0.420 \\
5 & 2711   & 00 34 27.83 & -52 22 23.1 & 39.24 & 6499 & 4.22 & +0.06 & 5.563 & -0.017 & 0.464 & 5.726 &  1.138 & 0.260 \\
6 & 3182   & 00 40 32.24 & -29 52 05.8 & 16.76 & 5643 & 4.32 & -0.29 & 8.695 &  0.084 & 0.627 & 8.963 &  1.297 & 0.438 \\
7 & 3479   & 00 44 26.65 & -26 30 56.4 & 30.89 & 5563 & 4.40 & -0.26 & 7.777 &  0.155 & 0.663 & 8.068 &  1.381 & 0.477 \\
8 & 3497   & 00 44 39.27 & -65 38 58.3 & 45.34 & 5638 & 4.41 & -0.32 & 6.538 &  0.106 & 0.652 & 6.822 &  1.328 & 0.465 \\
9 & 3704   & 00 47 30.75 & -36 56 24.7 & 20.51 & 6078 & 4.40 & -0.35 & 7.819 & -0.044 & 0.536 & 8.028 &  1.137 & 0.338 \\
10 & 3909  & 00 50 07.59 & -10 38 39.6 & 63.48 & 6352 & 4.45 & -0.03 & 5.178 & -0.005 & 0.507 & 5.368 &  1.166 & 0.307 \\
11 & 4544  & 00 58 11.69 & +80 06 49.3 &  9.23 & 5832 & 4.51 & -0.87 & 9.630 & -0.135 & 0.492 & 9.811 &  1.030 & 0.290 \\
12 & 4892  & 01 02 49.72 & -37 18 58.2 & 16.48 & 5994 & 4.42 & -0.29 & 8.501 &  0.004 & 0.575 & 8.735 &  1.199 & 0.381 \\
13 & 5163  & 01 06 05.15 & +01 42 23.1 & 10.84 & 5547 & 4.63 & -0.74 & 9.416 & -0.007 & 0.584 & 9.656 &  1.191 & 0.391 \\
14 & 5301  & 01 07 48.66 & -08 14 01.3 & 18.23 & 5686 & 4.39 & -0.11 & 8.384 &  0.157 & 0.632 & 8.655 &  1.372 & 0.443 \\
15 & 5862  & 01 15 11.12 & -45 31 54.0 & 66.16 & 6145 & 4.24 & +0.17 & 4.954 &  0.097 & 0.570 & 5.185 &  1.290 & 0.376 \\
16 & 6159  & 01 18 59.99 & -08 56 22.2 & 15.35 & 5653 & 4.56 & -0.67 & 8.862 & -0.027 & 0.583 & 9.101 &  1.171 & 0.390 \\
17 & 7217  & 01 32 57.60 & +23 41 44.0 & 14.60 & 5550 & 4.20 & -0.48 & 8.960 &  0.043 & 0.598 & 9.209 &  1.246 & 0.406 \\
18 & 7459  & 01 36 05.80 & -61 05 03.0 & 10.87 & 5759 & 4.31 & -1.23 & 10.07 & -0.167 & 0.520 &10.269 &  1.008 & 0.321 \\
19 & 7978  & 01 42 29.32 & -53 44 27.0 & 57.36 & 6219 & 4.41 & +0.05 & 5.531 & -0.006 & 0.527 & 5.734 &  1.172 & 0.329 \\
20 & 8859  & 01 53 57.68 & +10 36 50.5 & 24.12 & 6420 & 4.21 & -0.30 & 6.730 & -0.062 & 0.444 & 6.880 &  1.086 & 0.238 \\
21 & 9085  & 01 56 59.98 & -51 45 58.5 & 37.22 & 6334 & 4.29 & -0.26 & 6.088 & -0.062 & 0.477 & 6.259 &  1.098 & 0.274 \\
22 & 10449 & 02 14 40.30 & -01 12 05.1 & 15.87 & 5566 & 4.64 & -0.87 & 9.055 & -0.077 & 0.571 & 9.287 &  1.116 & 0.377 \\
23 & 10652 & 02 17 07.14 & +21 34 00.5 & 15.93 & 5499 & 4.58 & -0.67 & 8.954 & -0.008 & 0.592 & 9.199 &  1.193 & 0.400 \\
24 & 12294 & 02 38 21.50 & +02 26 44.4 &  5.82 & 6069 & 4.58 & -0.95 & 10.47 & -0.171 & 0.443 &10.619 &  0.977 & 0.237 \\
25 & 12306 & 02 38 27.86 & +30 48 59.9 & 29.17 & 5832 & 4.44 & -0.52 & 7.306 & -0.015 & 0.569 & 7.536 &  1.178 & 0.374 \\
26 & 12444 & 02 40 12.42 & -09 27 10.4 & 45.96 & 6383 & 4.41 & +0.09 & 5.766 & -0.013 & 0.517 & 5.963 &  1.161 & 0.318 \\
27 & 12653 & 02 42 33.47 & -50 48 01.1 & 58.25 & 6375 & 4.73 & +0.27 & 5.394 &  0.073 & 0.553 & 5.614 &  1.260 & 0.357 \\
28 & 12777 & 02 44 12.00 & +49 13 42.0 & 89.87 & 5309 & 4.30 & -0.02 & 4.090 & -0.009 & 0.482 & 4.264 &  1.153 & 0.279 \\
29 & 12889 & 02 45 41.01 & -38 09 31.4 & 20.16 & 5999 & 4.23 & -0.13 & 7.625 &  0.043 & 0.576 & 7.860 &  1.238 & 0.382 \\
30 & 13366 & 02 51 58.36 & +11 22 11.9 & 16.39 & 5856 & 4.26 & -0.69 & 8.234 & -0.096 & 0.499 & 8.419 &  1.072 & 0.298 \\
31 & 14241 & 03 03 38.96 & -05 39 58.7 & 28.54 & 5430 & 4.33 & -0.45 & 8.048 &  0.108 & 0.658 & 8.335 &  1.333 & 0.472 \\
32 & 15131 & 03 15 06.39 & -45 39 53.4 & 41.34 & 5985 & 4.54 & -0.43 & 6.748 & -0.023 & 0.576 & 6.983 &  1.172 & 0.382 \\
33 & 15158 & 03 15 22.52 & -01 10 43.1 & 10.55 & 6191 & 4.18 & -0.06 & 8.460 &  0.009 & 0.501 & 8.647 &  1.177 & 0.300 \\
34 & 15381 & 03 18 19.98 & +18 10 17.8 & 20.16 & 5882 & 4.24 & +0.08 & 7.472 &  0.154 & 0.598 & 7.721 &  1.357 & 0.406 \\
35 & 16169 & 03 28 21.08 & -06 31 51.3 & 21.38 & 5650 & 4.34 & -0.57 & 8.210 &  0.022 & 0.612 & 8.468 &  1.230 & 0.422 \\
36 & 17147 & 03 40 22.06 & -03 13 01.1 & 39.12 & 5970 & 4.52 & -0.81 & 6.664 & -0.092 & 0.532 & 6.871 &  1.087 & 0.334 \\
37 & 22632 & 04 52 09.91 & -27 03 51.0 & 15.00 & 5909 & 4.22 & -1.62 & 9.106 & -0.196 & 0.481 & 9.280 &  0.965 & 0.278 \\
38 & 23555 & 05 03 53.95 & -41 44 41.8 & 31.50 & 6466 & 4.64 & +0.22 & 6.296 &  0.039 & 0.528 & 6.500 &  1.217 & 0.330 \\
39 & 23688 & 05 05 28.70 & +40 15 26.0 &  8.79 & 6293 & 4.41 & -0.89 & 9.647 & -0.163 & 0.416 & 9.779 &  0.975 & 0.207 \\
40 & 24030 & 05 09 56.96 & +05 33 26.7 &  8.66 & 5738 & 4.64 & -1.00 & 9.648 & -0.147 & 0.497 & 9.832 &  1.020 & 0.296 \\
41 & 25860 & 05 31 13.78 & +15 46 24.4 & 20.07 & 5543 & 4.56 & -0.35 & 8.506 &  0.088 & 0.627 & 8.774 &  1.301 & 0.438 \\
42 & 26617 & 05 39 27.44 & +03 57 02.7 &  8.79 & 5429 & 4.67 & -0.75 &10.116 & -0.025 & 0.554 &10.337 &  1.162 & 0.358 \\
43 & 27072 & 05 44 27.79 & -22 26 54.2 &112.02 & 6323 & 4.16 & -0.02 & 3.587 & -0.008 & 0.480 & 3.760 &  1.153 & 0.277 \\
44 & 28403 & 05 59 55.79 & -37 03 24.2 & 19.64 & 5668 & 4.31 & -0.33 & 8.575 &  0.046 & 0.620 & 8.838 &  1.257 & 0.430 \\
45 & 28671 & 06 03 14.86 & +19 21 38.7 & 16.81 & 5396 & 4.39 & -1.11 & 9.260 & -0.055 & 0.602 & 9.511 &  1.150 & 0.411 \\
46 & 29432 & 06 12 00.60 & +06 46 59.0 & 42.55 & 5726 & 4.58 & -0.16 & 6.826 &  0.119 & 0.625 & 7.092 &  1.332 & 0.436 \\
47 & 31188 & 06 32 37.99 & -06 29 18.5 & 16.79 & 5789 & 4.54 & -0.59 & 8.514 & -0.094 & 0.525 & 8.716 &  1.083 & 0.326 \\
48 & 33582 & 06 58 38.54 & -00 28 49.7 & 12.56 & 5773 & 4.11 & -0.62 & 8.987 & -0.024 & 0.569 & 9.217 &  1.169 & 0.374 \\
49 & 34017 & 07 03 30.46 & +29 20 13.5 & 52.27 & 5920 & 4.42 & -0.12 & 5.921 &  0.058 & 0.588 & 6.163 &  1.258 & 0.395 \\
50 & 34511 & 07 09 04.96 & +15 25 17.7 & 21.64 & 5789 & 4.44 & -0.11 & 7.981 &  0.116 & 0.614 & 8.240 &  1.325 & 0.424 \\
51 & 35139 & 07 15 50.80 & -13 02 58.1 & 30.95 & 5771 & 4.47 & -0.52 & 7.749 & -0.018 & 0.605 & 8.002 &  1.188 & 0.414 \\
52 & 36491 & 07 30 29.02 & +18 57 40.6 & 20.20 & 5869 & 4.31 & -0.95 & 8.482 & -0.120 & 0.514 & 8.677 &  1.053 & 0.314 \\
53 & 36849 & 07 34 35.11 & +16 54 04.0 & 12.37 & 6012 & 4.16 & -0.84 & 8.922 & -0.116 & 0.501 & 9.109 &  1.052 & 0.300 \\
54 & 37419 & 07 40 54.38 & -26 21 48.6 & 19.16 & 5715 & 4.38 & -0.33 & 8.655 &  0.064 & 0.616 & 8.915 &  1.274 & 0.426 \\
55 & 38541 & 07 53 33.12 & +30 36 18.3 & 34.30 & 5316 & 4.60 & -1.84 & 8.288 & -0.128 & 0.615 & 8.548 &  1.081 & 0.425 \\
56 & 38769 & 07 56 10.20 & +50 32 27.3 & 11.56 & 5726 & 4.15 & -0.79 & 8.780 & -0.072 & 0.504 & 8.969 &  1.097 & 0.303 \\
57 & 40118 & 08 11 38.64 & +32 27 25.7 & 45.90 & 5484 & 4.58 & -0.47 & 6.817 &  0.125 & 0.676 & 7.116 &  1.356 & 0.492 \\
58 & 40778 & 08 19 22.60 & +54 05 10.0 & 10.35 & 6006 & 4.23 & -1.55 & 9.699 & -0.222 & 0.467 & 9.864 &  0.934 & 0.263 \\
59 & 41484 & 08 27 36.80 & +45 39 11.0 & 44.94 & 5703 & 4.46 & -0.08 & 6.312 &  0.118 & 0.623 & 6.577 &  1.330 & 0.434 \\
60 & 42356 & 08 38 08.51 & +26 02 56.3 & 22.91 & 5962 & 4.42 & +0.17 & 7.560 &  0.131 & 0.626 & 7.827 &  1.344 & 0.437 \\
61 & 43595 & 08 52 44.51 & +22 33 30.2 &  7.87 & 5506 & 4.67 & -0.84 &10.789 & -0.049 & 0.587 &11.031 &  1.150 & 0.394 \\
62 & 43726 & 08 54 17.90 & -05 26 04.0 & 57.52 & 5763 & 4.37 & +0.01 & 6.007 &  0.211 & 0.665 & 6.299 &  1.438 & 0.480 \\
63 & 44116 & 08 59 06.00 & -00 37 26.0 & 12.70 & 6275 & 4.10 & -0.58 & 8.456 & -0.109 & 0.432 & 8.598 &  1.035 & 0.225 \\
64 & 44713 & 09 06 38.83 & -43 29 31.1 & 26.83 & 5823 & 4.34 & +0.14 & 7.261 &  0.191 & 0.651 & 7.544 &  1.413 & 0.464 \\
65 & 44811 & 09 07 56.60 & -50 28 57.0 & 24.53 & 5824 & 4.45 & -0.64 & 7.622 & -0.063 & 0.532 & 7.829 &  1.116 & 0.334 \\
66 & 47048 & 09 35 16.70 & -49 07 48.9 & 10.26 & 6630 & 4.12 & -0.50 & 8.451 & -0.118 & 0.366 & 8.551 &  1.002 & 0.152 \\
67 & 47174 & 09 36 49.50 & +57 54 41.0 & 11.99 & 5603 & 4.33 & -0.45 & 9.940 &  0.025 & 0.623 &10.205 &  1.237 & 0.434 \\
68 & 48209 & 09 49 42.82 & +65 18 15.0 & 13.10 & 5415 & 4.64 & -0.65 & 9.601 &  0.078 & 0.643 & 9.879 &  1.297 & 0.455 \\
69 & 49285 & 10 03 37.38 & -29 02 36.0 & 21.13 & 5520 & 4.29 & -0.21 & 8.073 &  0.191 & 0.668 & 8.367 &  1.419 & 0.483 \\
70 & 49615 & 10 07 33.80 & -06 26 21.0 & 21.25 & 6019 & 4.29 & -0.43 & 7.703 & -0.064 & 0.504 & 7.892 &  1.105 & 0.303 \\
71 & 49793 & 10 09 49.58 & -36 45 14.9 & 22.58 & 5764 & 4.33 & -0.63 & 8.054 & -0.026 & 0.587 & 8.296 &  1.173 & 0.394 \\
72 & 49942 & 10 11 48.07 & +23 45 18.7 & 16.86 & 5676 & 4.36 & -0.27 & 8.397 &  0.110 & 0.625 & 8.663 &  1.323 & 0.436 \\
73 & 50834 & 10 22 46.93 & -45 28 14.2 &  8.09 & 6552 & 4.23 & -0.27 & 9.239 & -0.089 & 0.412 & 9.369 &  1.047 & 0.203 \\
74 & 50965 & 10 24 35.68 & -05 31 10.8 &  9.14 & 5715 & 4.56 & -0.57 & 9.753 & -0.011 & 0.555 & 9.974 &  1.177 & 0.359 \\
75 & 52673 & 10 46 14.24 & -29 20 25.5 & 14.07 & 5541 & 4.62 & -0.66 & 9.531 &  0.021 & 0.627 & 9.799 &  1.234 & 0.438 \\
76 & 52771 & 10 47 23.16 & +28 23 55.9 & 10.45 & 5354 & 4.77 & -1.98 &10.203 & -0.232 & 0.488 &10.381 &  0.932 & 0.286 \\
77 & 53537 & 10 57 09.60 & +21 48 17.0 & 20.32 & 6019 & 4.43 & +0.05 & 7.928 &  0.117 & 0.616 & 8.188 &  1.327 & 0.426 \\
78 & 53721 & 10 59 28.00 & +40 25 49.0 & 71.11 & 5882 & 4.34 & +0.01 & 5.027 &  0.123 & 0.605 & 5.280 &  1.329 & 0.414 \\
79 & 54469 & 11 08 40.07 & -44 15 33.7 &  9.27 & 6098 & 4.34 & -0.43 & 9.749 & -0.035 & 0.542 & 9.962 &  1.148 & 0.345 \\
80 & 54641 & 11 11 00.74 & -65 25 37.8 & 18.36 & 6168 & 4.29 & -1.14 & 8.083 & -0.176 & 0.458 & 8.242 &  0.977 & 0.253 \\
\hline
\end{tabular}
\end{center}
}
\end{table*}

\begin{table*}
\setcounter{table}{1}
\setlength{\tabcolsep}{1pt} 
{\scriptsize
\begin{center}
\begin{tabular}{cccccccccccccc}
\hline
ID & Hip & $\alpha_{2000}$ &$\delta_{2000}$ & $\pi$  & $T_{eff}$ & $\log g$ & [Fe/H] & $V_o$ & $(U-B)_o$ & $(B-V)_o$ & $g_o$ &  $(u-g)_o$ & $(g-r)_o$\\
   &     & (hh:mm:ss.ss) & (dd:mm:ss.s) & (mas)  & (K) & (cgs)& (dex) & (mag) & (mag) & (mag) & (mag) & (mag) & (mag)\\
\hline
81 & 54924 & 11 14 49.93 & -23 38 47.9  & 15.02  & 5682 & 4.34 & -0.80 & 9.018 & -0.008 & 0.568 & 9.248 &  1.184 & 0.373 \\
82 & 55592 & 11 23 16.23 & +19 53 37.7  &  8.35  & 5995 & 4.14 & -1.02 & 9.942 & -0.147 & 0.470 &10.109 &  1.010 & 0.266 \\
83 & 55761 & 11 25 31.80 & +42 37 58.0  & 22.95  & 5741 & 4.35 & -0.62 & 7.847 & -0.043 & 0.536 & 8.056 &  1.138 & 0.338 \\
84 & 56664 & 11 37 08.12 & -39 28 12.0  &  9.45  & 6364 & 4.07 & -0.68 & 8.636 & -0.109 & 0.412 & 8.766 &  1.027 & 0.203 \\
85 & 57017 & 11 41 22.48 & -26 40 01.9  & 18.78  & 6380 & 4.35 & -0.43 & 7.508 & -0.065 & 0.463 & 7.670 &  1.090 & 0.259 \\
86 & 57265 & 11 44 35.70 & +25 32 12.0  &  6.25  & 5928 & 4.23 & -0.93 &10.318 & -0.143 & 0.462 & 10.48 &  1.011 & 0.257 \\
87 & 57443 & 11 46 31.10 & -40 30 01.0  &108.45  & 5524 & 4.29 & -0.35 & 4.879 &  0.098 & 0.662 & 5.169 &  1.324 & 0.476 \\
88 & 57450 & 11 46 35.15 & +50 52 54.7  & 12.85  & 5315 & 4.74 & -1.50 & 9.902 & -0.154 & 0.555 &10.123 &  1.034 & 0.359 \\
89 & 58843 & 12 04 05.56 & +03 20 26.7  & 14.24  & 5726 & 4.41 & -0.84 & 9.194 & -0.059 & 0.579 & 9.431 &  1.137 & 0.385 \\
90 & 58950 & 12 05 13.41 & -28 43 02.0  & 27.85  & 5686 & 4.45 & -0.19 & 7.757 &  0.134 & 0.642 & 8.034 &  1.353 & 0.454 \\
91 & 59380 & 12 10 57.93 & -46 19 19.1  & 27.31  & 5901 & 4.38 & -0.58 & 7.502 & -0.059 & 0.554 & 7.723 &  1.128 & 0.358 \\
92 & 61053 & 12 30 50.10 & +53 04 36.0  & 45.92  & 6060 & 4.35 & -0.11 & 6.193 &  0.013 & 0.545 & 6.408 &  1.197 & 0.348 \\
93 & 61619 & 12 37 39.13 & +38 41 03.7  & 17.81  & 6429 & 4.26 & +0.04 & 7.338 & -0.023 & 0.476 & 7.509 &  1.136 & 0.273 \\
94 & 62108 & 12 43 43.22 & -44 40 31.6  &  8.17  & 6293 & 4.37 & -1.53 & 9.841 & -0.176 & 0.441 & 9.989 &  0.971 & 0.234 \\
95 & 62207 & 12 44 59.40 & +39 16 44.0  & 57.55  & 5795 & 4.15 & -0.59 & 5.945 & -0.045 & 0.546 & 6.161 &  1.139 & 0.349 \\
96 & 62809 & 12 52 11.64 & -56 34 28.0  & 20.67  & 5658 & 4.38 & -0.77 & 8.424 & -0.032 & 0.596 & 8.672 &  1.170 & 0.404 \\
97 & 64345 & 13 11 21.40 & +09 37 33.5  & 16.88  & 5598 & 4.37 & -0.60 & 8.718 &  0.047 & 0.613 & 8.977 &  1.255 & 0.423 \\
98 & 64394 & 13 11 52.40 & +27 52 41.0  &109.54  & 6029 & 4.38 & +0.03 & 4.228 &  0.067 & 0.570 & 4.459 &  1.260 & 0.376 \\
99 & 64698 & 13 15 36.97 & +09 00 57.7  & 18.48  & 5624 & 4.35 & -0.13 & 8.425 &  0.164 & 0.652 & 8.709 &  1.386 & 0.465 \\
100 & 64747 & 13 16 11.25 & +35 53 09.1 & 21.87  & 5718 & 4.47 & -0.21 & 8.271 &  0.118 & 0.637 & 8.545 &  1.335 & 0.449 \\
101 & 65418 & 13 24 30.60 & +20 27 22.0 &  3.12  & 6043 & 4.18 & -1.56 &12.138 & -0.186 & 0.433 &12.281 &  0.958 & 0.226 \\
102 & 66238 & 13 34 32.65 & -38 54 26.0 & 33.27  & 5660 & 4.42 & -0.22 & 7.279 &  0.117 & 0.665 & 7.571 &  1.344 & 0.480 \\
103 & 67655 & 13 51 40.40 & -57 26 08.0 & 39.42  & 5396 & 4.38 & -0.93 & 7.941 & -0.002 & 0.653 & 8.225 &  1.221 & 0.466 \\
104 & 67863 & 13 53 58.12 & -46 32 19.5 & 16.70  & 5692 & 4.35 & -0.76 & 9.004 & -0.055 & 0.593 & 9.250 &  1.146 & 0.401 \\
105 & 68030 & 13 55 50.00 & +14 03 23.0 & 40.22  & 6081 & 4.40 & -0.38 & 6.151 & -0.087 & 0.498 & 6.336 &  1.080 & 0.297 \\
106 & 70319 & 14 23 15.30 & +01 14 30.0 & 58.17  & 5597 & 4.44 & -0.41 & 6.242 &  0.081 & 0.634 & 6.514 &  1.297 & 0.446 \\
107 & 70681 & 14 27 24.91 & -18 24 40.4 & 21.04  & 5484 & 4.48 & -1.30 & 9.259 & -0.113 & 0.583 & 9.498 &  1.085 & 0.390 \\
108 & 70829 & 14 29 03.26 & -46 44 28.0 & 19.18  & 5471 & 4.47 & -0.67 & 8.903 &  0.084 & 0.673 & 9.200 &  1.314 & 0.488 \\
109 & 71076 & 14 32 04.50 & +18 50 10.0 & 17.24  & 6038 & 4.22 & -0.39 & 7.787 & -0.085 & 0.493 & 7.969 &  1.080 & 0.291 \\
110 & 71735 & 14 40 28.26 & -57 01 46.4 & 37.71  & 5513 & 4.43 & -0.35 & 7.348 &  0.141 & 0.658 & 7.635 &  1.366 & 0.472 \\
111 & 72407 & 14 48 18.70 & +58 54 36.1 & 10.65  & 5567 & 4.56 & -0.54 & 9.750 &  0.071 & 0.617 &10.011 &  1.281 & 0.427 \\
112 & 72673 & 14 51 31.72 & -60 55 50.8 & 20.06  & 6374 & 4.29 & -0.56 & 7.117 & -0.117 & 0.431 & 7.259 &  1.026 & 0.224 \\
113 & 74067 & 15 08 12.57 & -07 54 47.5 & 26.62  & 5695 & 4.38 & -0.81 & 7.969 & -0.057 & 0.580 & 8.206 &  1.140 & 0.387 \\
114 & 75181 & 15 21 48.15 & -48 19 03.5 & 67.51  & 5698 & 4.46 & -0.32 & 5.639 &  0.066 & 0.635 & 5.912 &  1.282 & 0.447 \\
115 & 78640 & 16 03 13.30 & +42 14 46.6 &  8.41  & 6069 & 4.30 & -1.43 & 9.835 & -0.202 & 0.468 &10.000 &  0.955 & 0.264 \\
116 & 81520 & 16 39 04.14 & -58 15 29.5 & 44.54  & 5746 & 4.62 & -0.44 & 7.002 &  0.025 & 0.609 & 7.258 &  1.232 & 0.418 \\
117 & 81681 & 16 41 08.21 & -02 51 26.2 & 33.82  & 5565 & 4.52 & -0.38 & 7.160 &  0.050 & 0.604 & 7.413 &  1.255 & 0.413 \\
118 & 83229 & 17 00 31.65 & -57 17 49.6 & 32.77  & 5865 & 4.41 & -0.49 & 6.987 & -0.024 & 0.568 & 7.217 &  1.168 & 0.373 \\
119 & 83489 & 17 03 49.15 & +17 11 21.1 & 13.43  & 5722 & 4.28 & -0.29 & 9.078 &  0.088 & 0.635 & 9.351 &  1.304 & 0.447 \\
120 & 84988 & 17 22 12.65 & -75 20 53.3 & 35.67  & 5686 & 4.25 & -0.70 & 6.977 & -0.016 & 0.593 & 7.223 &  1.185 & 0.401 \\
121 & 85007 & 17 22 27.65 & +24 52 46.0 & 34.12  & 6064 & 4.32 & -0.39 & 6.851 & -0.068 & 0.505 & 7.040 &  1.102 & 0.304 \\
122 & 85042 & 17 22 51.29 & -02 23 17.4 & 51.22  & 5648 & 4.46 &  0.00 & 6.217 &  0.220 & 0.659 & 6.505 &  1.445 & 0.473 \\
123 & 86013 & 17 34 43.06 & +06 00 51.6 & 19.38  & 5760 & 4.30 & -0.81 & 8.353 & -0.059 & 0.558 & 8.576 &  1.130 & 0.362 \\
124 & 86321 & 17 38 15.61 & +18 33 25.5 &  8.38  & 5760 & 4.59 & -0.87 & 9.724 & -0.117 & 0.465 & 9.888 &  1.038 & 0.261 \\
125 & 87062 & 17 47 28.00 & -08 46 48.0 &  9.59  & 6027 & 4.32 & -1.49 &10.266 & -0.209 & 0.485 &10.442 &  0.954 & 0.283 \\
126 & 88945 & 18 09 21.38 & +29 57 06.2 & 40.29  & 5869 & 4.62 & +0.02 & 6.838 &  0.087 & 0.616 & 7.098 &  1.297 & 0.426 \\
127 & 92270 & 18 48 16.40 & +23 30 53.1 & 34.78  & 6376 & 4.23 & -0.05 & 6.098 &  0.015 & 0.494 & 6.280 &  1.181 & 0.292 \\
128 & 92532 & 18 51 25.20 & +38 37 36.0 & 32.72  & 5825 & 4.30 & -0.56 & 7.138 & -0.048 & 0.536 & 7.347 &  1.133 & 0.338 \\
129 & 92781 & 18 54 23.20 & -04 36 18.6 & 14.59  & 5765 & 4.33 & -0.69 & 9.000 & -0.067 & 0.561 & 9.225 &  1.123 & 0.366 \\
130 & 93185 & 18 58 51.00 & +30 10 50.3 & 41.94  & 5864 & 4.54 & -0.29 & 6.775 & -0.004 & 0.575 & 7.009 &  1.191 & 0.381 \\
131 & 94129 & 19 09 39.24 & -21 28 10.8 & 17.60  & 5630 & 4.35 & -0.27 & 8.171 &  0.121 & 0.617 & 8.432 &  1.331 & 0.427 \\
132 & 94645 & 19 15 33.23 & -24 10 45.7 & 36.30  & 6365 & 4.56 & +0.24 & 6.232 &  0.065 & 0.533 & 6.439 &  1.245 & 0.335 \\
133 & 96124 & 19 32 40.33 & -28 01 11.3 & 36.72  & 5577 & 4.47 & -0.22 & 7.129 &  0.150 & 0.664 & 7.420 &  1.377 & 0.478 \\
134 & 96258 & 19 34 19.79 & +51 14 11.8 & 39.82  & 6380 & 4.15 & -0.03 & 5.721 & -0.012 & 0.464 & 5.884 &  1.143 & 0.260 \\
135 & 98355 & 19 58 58.54 & -46 05 17.0 & 19.11  & 6232 & 4.18 & -0.62 & 7.445 & -0.092 & 0.468 & 7.610 &  1.065 & 0.264 \\
136 & 99139 & 20 07 36.91 & -41 01 09.6 & 17.95  & 5612 & 4.32 & -0.37 & 8.809 &  0.093 & 0.630 & 9.078 &  1.308 & 0.441 \\
137 & 99938 & 20 16 38.08 & -07 26 37.8 & 17.64  & 5732 & 4.48 & -0.54 & 8.360 & -0.031 & 0.566 & 8.588 &  1.161 & 0.371 \\
138 & 100017 & 20 17 31.30 & +66 51 13.0 & 56.92 & 5782 & 4.44 & -0.19 & 5.854 &  0.045 & 0.567 & 6.083 &  1.237 & 0.372 \\
139 & 100279 & 20 20 24.60 & +06 01 53.0 & 10.46 & 5673 & 4.31 & -0.72 &10.039 & -0.070 & 0.583 &10.278 &  1.128 & 0.390 \\
140 & 100568 & 20 23 35.85 & -21 22 14.2 & 22.78 & 5800 & 4.48 & -1.10 & 8.628 & -0.139 & 0.541 & 8.840 &  1.044 & 0.344 \\
141 & 100792 & 20 26 11.92 & +09 27 00.4 & 17.00 & 6002 & 4.31 & -1.11 & 8.310 & -0.182 & 0.472 & 8.478 &  0.976 & 0.268 \\
142 & 102018 & 20 40 22.33 & -24 07 04.9 & 24.89 & 5933 & 4.12 & +0.13 & 7.187 &  0.137 & 0.594 & 7.433 &  1.339 & 0.402 \\
143 & 102046 & 20 40 49.38 & -18 47 33.3 & 16.15 & 6009 & 4.13 & -1.05 & 8.204 & -0.137 & 0.486 & 8.381 &  1.026 & 0.284 \\
144 & 102762 & 20 49 15.19 & -20 37 50.8 & 17.11 & 5948 & 4.30 & -0.04 & 8.070 &  0.067 & 0.575 & 8.304 &  1.262 & 0.381 \\
145 & 102805 & 20 49 37.80 & +12 32 42.0 & 32.66 & 6337 & 4.31 & -0.32 & 5.990 & -0.095 & 0.414 & 6.121 &  1.042 & 0.205 \\
146 & 103458 & 20 57 40.07 & -44 07 45.7 & 45.17 & 5843 & 4.53 & -0.61 & 6.507 & -0.045 & 0.582 & 6.746 &  1.152 & 0.389 \\
147 & 103498 & 20 58 08.52 & -48 12 13.5 & 18.95 & 5856 & 4.23 & -1.11 & 8.271 & -0.134 & 0.514 & 8.466 &  1.039 & 0.314 \\
148 & 103897 & 21 03 06.10 & +29 28 56.0 &  7.70 & 5607 & 4.39 & -0.67 &10.083 &  0.005 & 0.576 &10.318 &  1.200 & 0.382 \\
149 & 104659 & 21 11 59.03 & +17 43 39.9 & 29.10 & 5973 & 4.35 & -1.08 & 7.343 & -0.164 & 0.502 & 7.530 &  1.005 & 0.301 \\
150 & 107294 & 21 43 57.12 & +27 23 24.0 &  9.03 & 5929 & 4.63 & -1.14 & 9.993 & -0.153 & 0.462 &10.155 &  1.001 & 0.257 \\
151 & 107877 & 21 51 24.61 & -23 16 14.2 & 24.91 & 6355 & 4.20 & -0.24 & 6.850 & -0.049 & 0.477 & 7.021 &  1.111 & 0.274 \\
152 & 108468 & 21 58 24.32 & -12 39 52.8 & 29.93 & 5799 & 4.25 & -0.08 & 7.188 &  0.110 & 0.618 & 7.450 &  1.320 & 0.428 \\
153 & 108736 & 22 01 36.52 & -53 05 36.9 & 27.95 & 5990 & 4.47 & -0.29 & 7.088 &  0.022 & 0.565 & 7.316 &  1.213 & 0.370 \\
154 & 109144 & 22 06 33.17 & +01 51 25.7 & 19.52 & 6272 & 4.21 & -0.08 & 7.216 &  0.014 & 0.515 & 7.412 &  1.187 & 0.315 \\
155 & 109381 & 22 09 34.61 & -41 13 29.6 & 23.53 & 5803 & 4.46 & +0.13 & 7.828 &  0.208 & 0.657 & 8.115 &  1.432 & 0.471 \\
156 & 109646 & 22 12 43.50 & -06 28 08.0 & 27.64 & 5910 & 4.25 & -0.64 & 7.408 & -0.087 & 0.510 & 7.600 &  1.085 & 0.310 \\
157 & 110035 & 22 17 15.14 & +12 53 54.6 & 32.22 & 5907 & 4.42 & -0.14 & 7.010 &  0.069 & 0.590 & 7.254 &  1.269 & 0.397 \\
158 & 110341 & 22 20 55.80 & +08 11 12.3 & 32.33 & 6577 & 4.35 & -0.09 & 6.147 & -0.059 & 0.440 & 6.294 &  1.088 & 0.233 \\
159 & 110560 & 22 23 49.10 & +24 23 33.0 &  5.32 & 5791 & 4.14 & -0.53 &10.544 & -0.013 & 0.540 &10.756 &  1.169 & 0.343 \\
160 & 111565 & 22 36 07.70 & -54 36 38.2 & 31.82 & 5534 & 4.44 & -0.48 & 7.571 &  0.098 & 0.662 & 7.861 &  1.324 & 0.476 \\
161 & 112811 & 22 50 45.94 & +01 51 54.6 & 16.56 & 5347 & 4.64 & -0.70 & 9.283 &  0.069 & 0.664 & 9.574 &  1.296 & 0.478 \\
162 & 113688 & 23 01 33.17 & +19 16 10.7 & 12.07 & 5757 & 4.29 & -0.14 & 8.596 &  0.073 & 0.586 & 8.837 &  1.272 & 0.393 \\
163 & 113896 & 23 03 57.30 & -04 47 42.0 & 34.03 & 5872 & 4.28 & -0.18 & 6.658 &  0.047 & 0.573 & 6.891 &  1.241 & 0.379 \\
164 & 114450 & 23 10 43.49 & +18 54 32.6 & 14.48 & 5935 & 4.40 & -0.05 & 8.427 &  0.024 & 0.552 & 8.646 &  1.211 & 0.356 \\
165 & 114702 & 23 14 07.47 & -08 55 27.6 & 25.60 & 6028 & 4.33 & -0.34 & 7.529 & -0.043 & 0.546 & 7.745 &  1.141 & 0.349 \\
166 & 114924 & 23 16 42.30 & +53 12 49.0 & 48.77 & 6134 & 4.21 &  0.00 & 5.564 &  0.007 & 0.519 & 5.762 &  1.182 & 0.320 \\
167 & 116106 & 23 31 31.50 & -04 05 15.0 & 38.29 & 6005 & 4.42 & -0.24 & 6.480 & -0.031 & 0.520 & 6.679 &  1.144 & 0.321 \\
168 & 118115 & 23 57 33.52 & -09 38 51.1 & 20.84 & 5833 & 4.39 & +0.02 & 7.832 &  0.139 & 0.631 & 8.102 &  1.354 & 0.442 \\
\hline
\end{tabular}
\end{center}
}
\end{table*}

\begin{table*}
\setlength{\tabcolsep}{2pt} 
{\scriptsize
\caption{Photometric parameters for the stars in Hyades cluster. $M_V$ and $M_g$ are the absolute magnitudes in the $UBV$ and $ugr$ systems, respectively, the other symbols as in Table 1.} 
\begin{center}
\begin{tabular}{cccccccccc}
\hline
ID & Hip   & $V_o$  &$(U-B)_o$ & $(B-V)_o$ & $M_V$ & $g_o$ & $(u-g)_o$ & $(g-r)_o$ & $M_g$ \\
\hline
1  & 18327 &  8.990 & 0.640 & 0.890 & 5.880 &  9.426 & 1.948 & 0.726 & 6.314 \\
2  & 18946 & 10.130 & 1.030 & 1.070 & 6.960 & 10.682 & 2.402 & 0.923 & 7.516 \\
3  & 19098 &  9.290 & 0.580 & 0.890 & 5.630 &  9.726 & 1.888 & 0.726 & 6.067 \\
4  & 19148 &  7.850 & 0.100 & 0.600 & 4.540 &  8.100 & 1.304 & 0.408 & 4.789 \\
5  & 19207 & 10.480 & 1.120 & 1.180 & 7.400 & 11.103 & 2.531 & 1.043 & 8.021 \\
6  & 19263 &  9.940 & 0.890 & 0.990 & 6.410 & 10.441 & 2.233 & 0.835 & 6.910 \\
7  & 19316 & 11.270 & 1.260 & 1.320 & 8.340 & 11.982 & 2.722 & 1.196 & 9.051 \\
8  & 19504 &  6.620 &-0.010 & 0.420 & 3.390 &  6.755 & 1.129 & 0.211 & 3.527 \\
9  & 19781 &  8.460 & 0.240 & 0.700 & 5.130 &  8.774 & 1.480 & 0.518 & 5.448 \\
10 & 19786 &  8.060 & 0.170 & 0.640 & 4.900 &  8.336 & 1.388 & 0.452 & 5.172 \\
11 & 19793 &  8.090 & 0.200 & 0.660 & 4.710 &  8.379 & 1.425 & 0.474 & 4.998 \\
12 & 19796 &  7.120 & 0.050 & 0.510 & 3.850 &  7.312 & 1.222 & 0.310 & 4.044 \\
13 & 19808 & 10.750 & 1.180 & 1.180 & 7.880 & 11.373 & 2.591 & 1.043 & 8.505 \\
14 & 20082 &  9.600 & 0.800 & 0.990 & 5.980 & 10.101 & 2.143 & 0.835 & 6.483 \\
15 & 20146 &  8.450 & 0.310 & 0.730 & 5.120 &  8.784 & 1.560 & 0.551 & 5.451 \\
16 & 20349 &  6.800 &-0.030 & 0.430 & 3.210 &  6.941 & 1.113 & 0.222 & 3.351 \\
17 & 20350 &  6.810 & 0.000 & 0.440 & 3.320 &  6.957 & 1.147 & 0.233 & 3.469 \\
18 & 20357 &  6.610 & 0.000 & 0.410 & 3.010 &  6.738 & 1.136 & 0.201 & 3.142 \\
19 & 20480 &  8.850 & 0.350 & 0.760 & 5.340 &  9.203 & 1.611 & 0.583 & 5.695 \\
20 & 20491 &  7.180 & 0.030 & 0.450 & 3.650 &  7.334 & 1.180 & 0.244 & 3.800 \\
21 & 20492 &  9.120 & 0.540 & 0.860 & 5.790 &  9.537 & 1.837 & 0.693 & 6.202 \\
22 & 20527 & 10.900 & 1.260 & 1.280 & 7.530 & 11.587 & 2.707 & 1.152 & 8.221 \\
23 & 20557 &  7.140 & 0.040 & 0.520 & 3.910 &  7.339 & 1.215 & 0.321 & 4.109 \\
24 & 20567 &  6.970 & 0.000 & 0.440 & 3.330 &  7.117 & 1.147 & 0.233 & 3.475 \\
25 & 20741 &  8.120 & 0.200 & 0.660 & 4.810 &  8.409 & 1.425 & 0.474 & 5.094 \\
26 & 20762 & 10.440 & 1.040 & 1.150 & 7.220 & 11.043 & 2.441 & 1.010 & 7.826 \\
27 & 20815 &  7.420 & 0.060 & 0.540 & 4.060 &  7.632 & 1.242 & 0.343 & 4.271 \\
28 & 20826 &  7.510 & 0.050 & 0.550 & 4.180 &  7.728 & 1.236 & 0.354 & 4.393 \\
29 & 20850 &  9.040 & 0.530 & 0.840 & 5.740 &  9.444 & 1.820 & 0.671 & 6.142 \\
30 & 20899 &  7.850 & 0.130 & 0.610 & 4.510 &  8.107 & 1.337 & 0.419 & 4.770 \\
31 & 20978 &  9.110 & 0.560 & 0.860 & 6.060 &  9.527 & 1.857 & 0.693 & 6.473 \\
32 & 21066 &  7.040 & 0.000 & 0.470 & 3.790 &  7.207 & 1.157 & 0.266 & 3.955 \\
33 & 21099 &  8.600 & 0.350 & 0.740 & 5.290 &  8.940 & 1.604 & 0.562 & 5.633 \\
34 & 21112 &  7.770 & 0.060 & 0.540 & 4.180 &  7.982 & 1.242 & 0.343 & 4.387 \\
35 & 21152 &  6.400 & 0.000 & 0.410 & 3.210 &  6.528 & 1.136 & 0.201 & 3.336 \\
36 & 21256 & 10.670 & 1.130 & 1.240 & 7.810 & 11.331 & 2.563 & 1.109 & 8.467 \\
37 & 21261 & 10.740 & 1.170 & 1.190 & 7.340 & 11.369 & 2.585 & 1.054 & 7.969 \\
38 & 21267 &  6.620 &-0.010 & 0.430 & 3.370 &  6.761 & 1.133 & 0.222 & 3.511 \\
39 & 21637 &  7.530 & 0.100 & 0.580 & 4.290 &  7.767 & 1.297 & 0.387 & 4.524 \\
40 & 21723 & 10.040 & 1.000 & 1.080 & 6.880 & 10.598 & 2.376 & 0.934 & 7.438 \\
41 & 21788 &  8.830 & 0.240 & 0.700 & 5.120 &  9.144 & 1.480 & 0.518 & 5.430 \\
42 & 22271 & 10.600 & 1.100 & 1.180 & 7.360 & 11.223 & 2.511 & 1.043 & 7.986 \\
43 & 22380 &  8.970 & 0.510 & 0.820 & 5.710 &  9.361 & 1.793 & 0.649 & 6.101 \\
44 & 22422 &  7.740 & 0.110 & 0.580 & 4.300 &  7.977 & 1.307 & 0.387 & 4.533 \\
45 & 23069 &  8.890 & 0.350 & 0.740 & 5.310 &  9.230 & 1.604 & 0.562 & 5.649 \\
\hline
\end{tabular}
\end{center}
}
\end{table*}

\begin{figure}[p]
\begin{center}
\includegraphics[scale=0.8, angle=0]{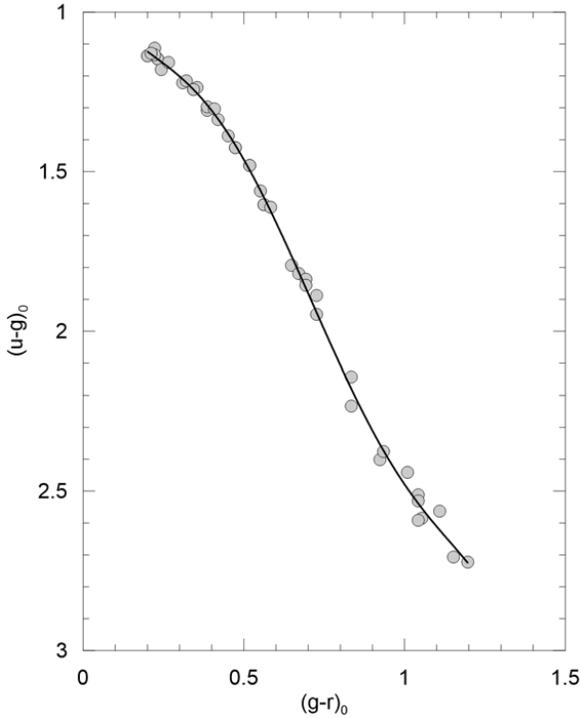}
\caption[] {The two-colour diagram of the Hyades cluster.}
\end{center}
\end{figure}

\begin{figure}[p]
\begin{center}
\includegraphics[scale=1.1, angle=0]{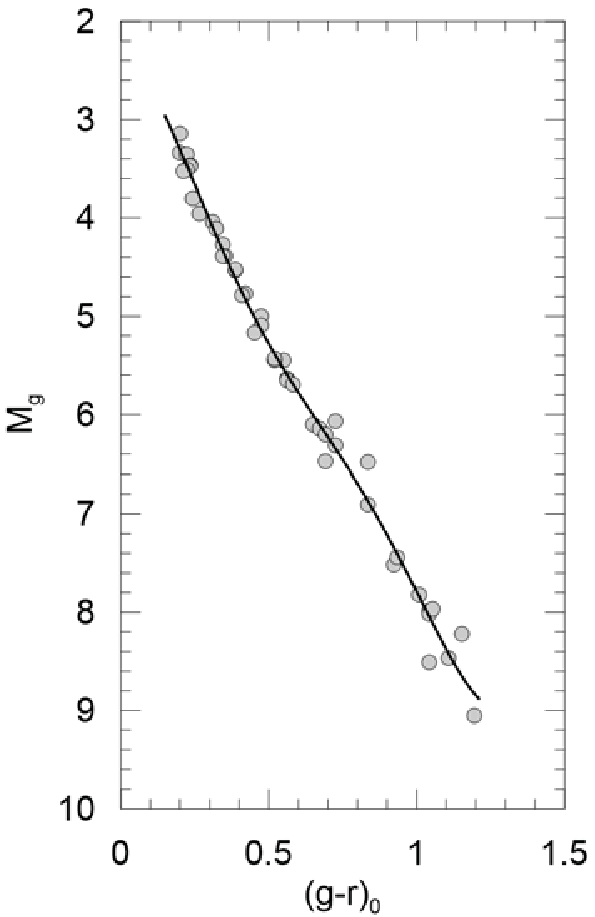}
\caption[] {The absolute magnitude-colour diagram of the Hyades cluster.}
\end{center}
\end{figure}

The two-colour diagram $(u-g)_0\times(g-r)_0$ and the absolute magnitude-colour diagram $M_g\times(g-r)_0$ of the Hyades cluster used in our calibrations are provided from \citet{Karaali03a} and \citet{Sandage69}, respectively, by transformation of their two-colour and absolute magnitude-colour diagrams from the $UBV$ system to the $ugr$ one. We should emphasize that the $M_V$ absolute magnitudes had been updated by considering the re-reduced {\it Hipparcos} \citep{vanLeeuwen07} trigonometric parallaxes before the necessary transformations. The results are given in Table 2, Fig. 4 and Fig. 5. The equations of the $(u-g)_0$ colour index and $M_g$ absolute magnitude for the Hyades cluster in terms of $(g-r)_0$ colour index are as follows:

\begin{eqnarray}
(u-g)_0=0.839+2.756(g-r)_0-10.744(g-r)_0^2\\ \nonumber
+24.252(g-r)_0^3-20.585(g-r)_0^4+5.961(g-r)_0^5.
\end{eqnarray}

\begin{eqnarray}
M_g=2.343+0.575(g-r)_0+32.996(g-r)_0^2\\ \nonumber
-71.036(g-r)_0^3+62.039(g-r)_0^4-19.128(g-r)_0^5.		
\end{eqnarray}

The squared correlation coefficients and the standard deviations of Eqs. (10) and (11) are $R^2=0.998$, $\sigma=0.027$ and $R^2=0.991$, $\sigma=0.164$, respectively. The range of the $(g-r)_0$ colour index for the Hyades cluster is $0.2<(g-r)_0 <1.2$ mag.

\begin{figure}[t]
\begin{center}
\includegraphics[scale=0.8, angle=0]{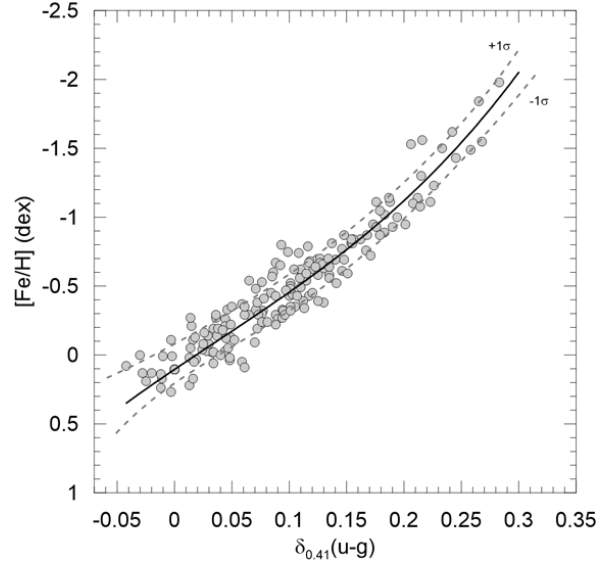}
\caption[] {Calibration of the iron abundance [Fe/H] in terms of reduced UV excess, $\delta_{0.41}$. The symbol $\sigma$ indicates the standard deviation estimated for the third order polynomial fitted for the calibration.}
\end{center}
\end{figure}

\section{Procedure}
\subsection{Calibration of [Fe/H] in terms of $\delta_{0.41}$}
We used the same procedure in Paper I for metallicity calibration. We evaluated the ultra-violet excess $\delta=(u-g)_H-(u-g)_*$ for each star, where $(u-g)_*$ and $(u-g)_H$ are the de-reddened UV indices for a sample star and the Hyades star with the same $(g-r)_0$ colour index. Then, we reduced it to the UV excess corresponding to the colour index $(g-r)_0=0.41$ mag via the appropriate guillotine factor ($f$, normalized factor) in \citet{Sandage69}, i.e. $\delta_{0.41}=f\times\delta$. Here, $(g-r)_0=0.41$ mag corresponds to $(B-V)_0=0.60$ mag in the $UBV$ system \citep[see also,][]{Karaali11}. Finally, we fitted the iron abundances [Fe/H] of the sample stars to a third order polynomial in terms of their reduced UV excesses, $\delta_{0.41}$ as follows (Fig. 6):

\begin{eqnarray}         
[Fe/H]=0.105(0.010)-5.633(0.521)\times \delta_{0.41}\\ \nonumber
+2.984(1.895)\times \delta_{0.41}^2-27.209(16.359)\times \delta_{0.41}^3.\\ \nonumber
\end{eqnarray}

Eq. (12) is the metallicity calibration in terms of the UV excess defined in the $ugr$ system. The squared correlation coefficient and the standard deviation for our calibration are $R^2=0.949$ and $\sigma=0.107$ mag, respectively. This calibration covers the stars with iron abundance $-2<$[Fe/H]$<0.3$ dex, and the range of the reduced UV excess is $-0.03<\delta_{0.41}<0.28$ mag. 
 
We tested the accuracy of our calibration by calculating the iron abundances, [Fe/H]$_{cal}$, of the sample stars via replacing their reduced UV excesses, $\delta_{0.41}$, into  Eq. (12) and compared them with the original ones, [Fe/H]$_{org}$. The distribution of the residuals, $\Delta$[Fe/H]=[Fe/H]$_{org}$-[Fe/H]$_{cal}$, are plotted in Fig. 7b. Their mean and the corresponding standard deviation are $\langle$[Fe/H]$\rangle=0.00$ and $\sigma=0.137$ dex, respectively. Zero mean residual and small standard deviation as well as the diagonal trend between the original and calculated iron metallicities (Fig. 7a) are indications for accuracy of our calibration.   

\begin{figure}
\begin{center}
\includegraphics[scale=0.8, angle=0]{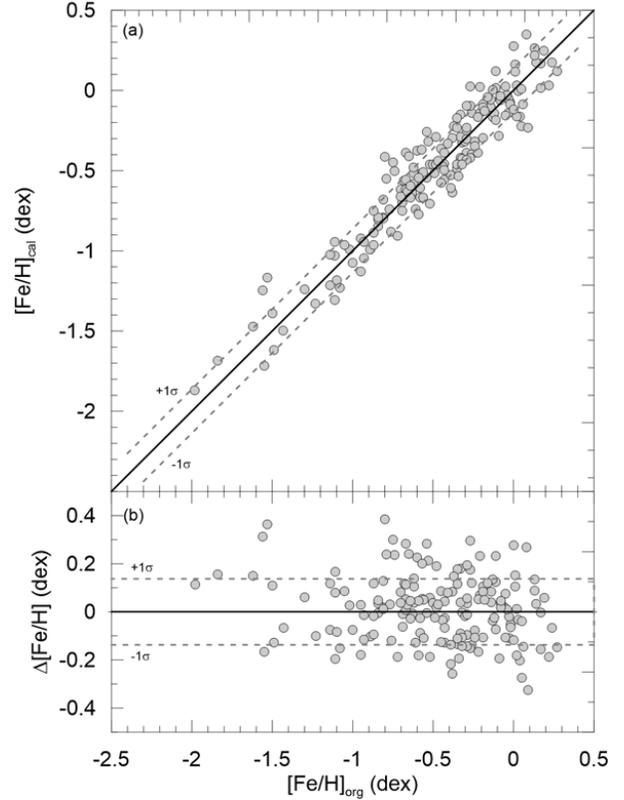}
\caption[] {The relation between the original iron abundances and the estimated ones (panel a), and the distribution of the residuals relative to the original iron abundances (panel b). The symbol $\sigma$ is the same as in Fig. 6.}
\end{center}
\end{figure}

\begin{figure}
\begin{center}
\includegraphics[scale=0.8, angle=0]{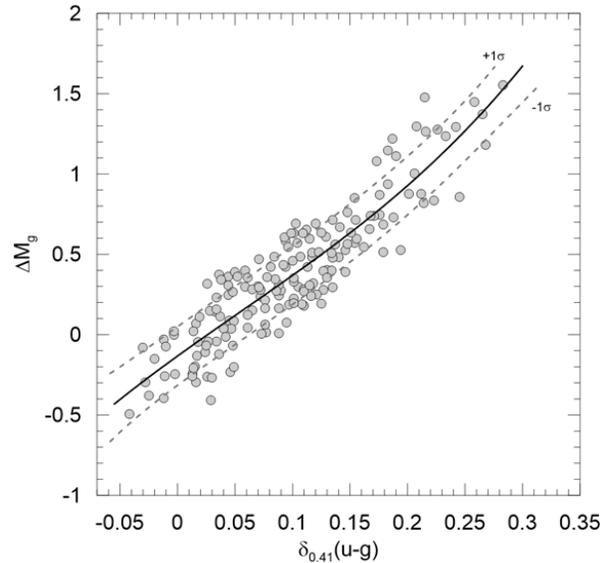}
\caption[] {Diagram for the absolute magnitude offset, $\Delta M_g$, versus UV excess, $\delta_{0.41}$. The symbol $\sigma$ is the same as in Fig. 6.}
\end{center}
\end{figure}

\subsection{Calibration of $\Delta M_g$ in terms of $\delta_{0.41}$}
Photometric parallax estimation is based on the calibration of the absolute magnitude offset in terms of reduced UV excess. We evaluated the difference between the absolute magnitude of each sample star, $(M_g)_*$, and the corresponding Hyades star, $(M_g)_H$, with the same $(g-r)_0$ colour index as follows:
\begin{equation}         
\Delta M_g=(M_g)_*-(M_g)_H.
\end{equation}
Then, we fitted the $\Delta M_g$ differences to a third order polynomial in terms of the reduced UV excesses of the sample stars as follows (Fig. 8):

\begin{eqnarray}         
\Delta M_g=-0.133(0.031)+5.169(0.965)\times \delta_{0.41}\\ \nonumber
-3.623(0.978)\times \delta_{0.41}^2+21.497(7.799)\times \delta_{0.41}^3.\\ \nonumber        
\end{eqnarray}

Eq. (14) is the absolute magnitude calibration in terms of reduced UV excess. Its squared correlation coefficient and standard deviation are $R^2=0.816$ and $\sigma=0.180$ mag, respectively. The range of the reduced UV excess is $-0.03<\delta_{0.41}<0.30$ dex and our calibration covers the absolute magnitude residuals $-0.5<\Delta M_g<1.6$ mag. 

We tested the accuracy of our calibration via the procedure explained in the following. We replaced the reduced UV excesses of the sample stars into Eq. (14) and evaluated the absolute magnitude offsets, $(\Delta M_g)_{cal}$, of the sample stars. Then, we adopted Eq. (13) and evaluated an absolute magnitude for each sample star:

\begin{equation}         
(M_g)_{cal}= (M_g)_H +(\Delta M_g)_{cal}.
\end{equation}
Here, $(M_g)_H$ is the absolute magnitude of the Hyades star with the same $(g-r)_0$ of the star in question. The $(M_g)_{cal}$ absolute magnitudes are tabulated in the seventh column of Table 3. Finally, we estimated the absolute magnitude residuals, the differences between the original and evaluated absolute magnitudes:
\begin{equation}         
Residuals=(M_g)_{org} - (M_g)_{cal}.
\end{equation}

The mean of the residuals ($\langle Res \rangle$) and the corresponding standard deviation ($\sigma$) are 0 and 0.179 mag, respectively. The evaluated absolute magnitudes and the residuals are plotted in Fig. 9. As in the case of metallicity calibration, zero mean residual, relatively small standard deviation and diagonal trend in the distribution of evaluated versus original absolute magnitudes is an indication for the accuracy of our absolute magnitude calibration.    

\begin{figure}[t]
\begin{center}
\includegraphics[scale=0.8, angle=0]{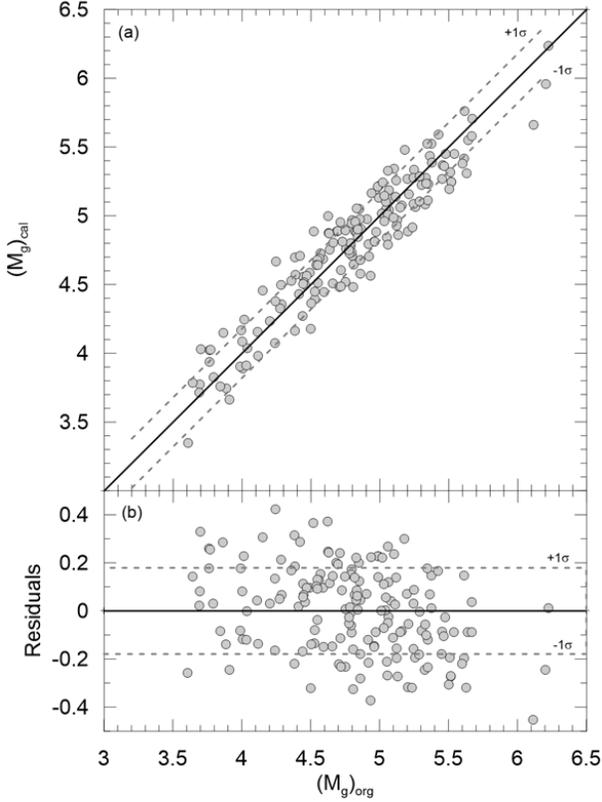}
\caption[] {The relation between the original absolute magnitudes  and the estimated ones (panel a), and the distribution of the residuals relative to the original absolute magnitudes (panel b). The symbol $\sigma$ is the same as in Fig. 6.}
\end{center}
\end{figure}

\begin{table*}
\setlength{\tabcolsep}{2pt} 
{\scriptsize
\caption{Apparent and absolute magnitudes for 168 sample stars used in the calibrations. The columns give: ID, Hipparcos number, trigonometric parallax ($\pi$), apparent magnitude ($g_0$), original absolute magnitude ($(M_g)_{org}$), absolute magnitude offset ($\Delta M_g$), estimated absolute magnitude $(M_g)_{cal}$ and the residuals.} 
\begin{center}
\begin{tabular}{cccccccc|cccccccc}
\hline
ID & Hip & $\pi$ & $g_0$ & $(M_g)_{org}$ & $\Delta M_g$ & $(M_g)_{cal}$ & Residual & ID & Hip & $\pi$ & $g_0$ & $(M_g)_{org}$ & $\Delta M_g$ & $(M_g)_{cal}$ & Residual \\
   &     & (mas) & (mag) & (mag) & (mag) & (mag) & (mag)&   &     & (mas) & (mag) & (mag) & (mag) & (mag) & (mag)\\
\hline
1 & 493 & 26.93 & 7.656 & 4.807 & 0.375 & 4.482 & -0.325 & 85 & 57017 & 18.78 & 7.670 & 4.038 & 0.310 & 4.037 & -0.001 \\
2 & 910 & 53.34 & 5.065 & 3.700 & -0.231 & 4.030 & 0.330 & 86 & 57265 & 6.25 & 10.480 & 4.459 & 0.745 & 4.496 & 0.037 \\
3 & 1599 &116.46 & 4.457 & 4.788 & 0.237 & 4.780 & -0.008 & 87 & 57443 & 108.45 & 5.169 & 5.345 & 0.196 & 5.523 & 0.178 \\
4 & 1976 & 21.27 & 7.805 & 4.444 & -0.379 & 4.558 & 0.114 & 88 & 57450 & 12.85 & 10.123 & 5.668 & 1.236 & 5.579 & -0.089 \\
5 & 2711 & 39.24 & 5.726 & 3.695 & -0.041 & 3.776 & 0.081 & 89 & 58843 & 14.24 & 9.431 & 5.199 & 0.596 & 5.270 & 0.071 \\
6 & 3182 & 16.76 & 8.963 & 5.084 & 0.153 & 5.125 & 0.041 & 90 & 58950 & 27.85 & 8.034 & 5.258 & 0.233 & 5.065 & -0.193 \\
7 & 3479 & 30.89 & 8.068 & 5.517 & 0.362 & 5.244 & -0.273 & 91 & 59380 & 27.31 & 7.723 & 4.905 & 0.480 & 4.977 & 0.072 \\
8 & 3497 & 45.34 & 6.822 & 5.104 & 0.016 & 5.341 & 0.237 & 92 & 61053 & 45.92 & 6.408 & 4.718 & 0.360 & 4.487 & -0.231 \\
9 & 3704 & 20.51 & 8.028 & 4.588 & 0.298 & 4.685 & 0.097 & 93 & 61619 & 17.81 & 7.509 & 3.762 & -0.068 & 3.939 & 0.177 \\
10 & 3909 & 63.48 & 5.368 & 4.381 & 0.308 & 4.162 & -0.219 & 94 & 62108 & 8.17 & 9.989 & 4.550 & 1.003 & 4.513 & -0.037 \\
11 & 4544 & 9.23 & 9.811 & 4.637 & 0.685 & 4.752 & 0.115 & 95 & 62207 & 57.55 & 6.161 & 4.961 & 0.596 & 4.811 & -0.150 \\
12 & 4892 & 16.48 & 8.735 & 4.820 & 0.243 & 4.886 & 0.066 & 96 & 62809 & 20.67 & 8.672 & 5.249 & 0.525 & 5.336 & 0.087 \\
13 & 5163 & 10.84 & 9.656 & 4.831 & 0.189 & 5.052 & 0.221 & 97 & 64345 & 16.88 & 8.977 & 5.114 & 0.273 & 5.140 & 0.026 \\
14 & 5301 & 18.23 & 8.655 & 4.959 & -0.002 & 4.813 & -0.146 & 98 & 64394 & 109.54 & 4.459 & 4.657 & 0.112 & 4.509 & -0.148 \\
15 & 5862 & 66.16 & 5.185 & 4.288 & -0.257 & 4.355 & 0.067 & 99 & 64698 & 18.48 & 8.709 & 5.043 & -0.045 & 5.047 & 0.004 \\
16 & 6159 & 15.35 & 9.101 & 5.032 & 0.397 & 5.144 & 0.112 & 100 & 64747 & 21.87 & 8.545 & 5.244 & 0.248 & 5.085 & -0.159 \\
17 & 7217 & 14.60 & 9.209 & 5.031 & 0.295 & 4.960 & -0.071 & 101 & 65418 & 3.12 & 12.281 & 4.752 & 1.263 & 4.520 & -0.232 \\
18 & 7459 & 10.87 & 10.269 & 5.450 & 1.278 & 5.270 & -0.180 & 102 & 66238 & 33.27 & 7.571 & 5.181 & 0.009 & 5.481 & 0.300 \\
19 & 7978 & 57.36 & 5.734 & 4.527 & 0.299 & 4.392 & -0.135 & 103 & 67655 & 39.42 & 8.225 & 6.204 & 1.111 & 5.959 & -0.245 \\
20 & 8859 & 24.12 & 6.880 & 3.792 & 0.216 & 3.824 & 0.032 & 104 & 67863 & 16.70 & 9.250 & 5.364 & 0.659 & 5.434 & 0.070 \\
21 & 9085 & 37.22 & 6.259 & 4.113 & 0.276 & 4.156 & 0.043 & 105 & 68030 & 40.22 & 6.336 & 4.358 & 0.356 & 4.527 & 0.169 \\
22 & 10449 & 15.87 & 9.287 & 5.290 & 0.739 & 5.286 & -0.004 & 106 & 70319 & 58.17 & 6.514 & 5.337 & 0.359 & 5.236 & -0.101 \\
23 & 10652 & 15.93 & 9.199 & 5.210 & 0.511 & 5.156 & -0.054 & 107 & 70681 & 21.04 & 9.498 & 6.113 & 1.478 & 5.660 & -0.453 \\
24 & 12294 & 5.82 & 10.619 & 4.444 & 0.875 & 4.503 & 0.059 & 108 & 70829 & 19.18 & 9.200 & 5.614 & 0.399 & 5.761 & 0.147 \\
25 & 12306 & 29.17 & 7.536 & 4.861 & 0.329 & 4.906 & 0.045 & 109 & 71076 & 17.24 & 7.969 & 4.152 & 0.193 & 4.458 & 0.306 \\
26 & 12444 & 45.96 & 5.963 & 4.275 & 0.124 & 4.325 & 0.050 & 110 & 71735 & 37.71 & 7.635 & 5.517 & 0.390 & 5.246 & -0.271 \\
27 & 12653 & 58.25 & 5.614 & 4.440 & 0.021 & 4.271 & -0.169 & 111 & 72407 & 10.65 & 10.011 & 5.148 & 0.282 & 5.060 & -0.088 \\
28 & 12777 & 89.87 & 4.264 & 4.032 & 0.159 & 3.913 & -0.119 & 112 & 72673 & 20.06 & 7.259 & 3.771 & 0.296 & 4.027 & 0.256 \\
29 & 12889 & 20.16 & 7.860 & 4.382 & -0.202 & 4.698 & 0.316 & 113 & 74067 & 26.62 & 8.206 & 5.332 & 0.716 & 5.277 & -0.055 \\
30 & 13366 & 16.39 & 8.419 & 4.492 & 0.483 & 4.588 & 0.096 & 114 & 75181 & 67.51 & 5.912 & 5.059 & 0.075 & 5.328 & 0.269 \\
31 & 14241 & 28.54 & 8.335 & 5.612 & 0.485 & 5.416 & -0.196 & 115 & 78640 & 8.41 & 10.000 & 4.624 & 0.859 & 4.997 & 0.373 \\
32 & 15131 & 41.34 & 6.983 & 5.065 & 0.481 & 5.041 & -0.024 & 116 & 81520 & 44.54 & 7.258 & 5.502 & 0.691 & 5.196 & -0.306 \\
33 & 15158 & 10.55 & 8.647 & 3.763 & -0.261 & 4.023 & 0.260 & 117 & 81681 & 33.82 & 7.413 & 5.059 & 0.279 & 5.009 & -0.050 \\
34 & 15381 & 20.16 & 7.721 & 4.243 & -0.493 & 4.378 & 0.135 & 118 & 83229 & 32.77 & 7.217 & 4.794 & 0.269 & 4.946 & 0.152 \\
35 & 16169 & 21.38 & 8.468 & 5.118 & 0.283 & 5.256 & 0.138 & 119 & 83489 & 13.43 & 9.351 & 4.991 & 0.007 & 5.218 & 0.227 \\
36 & 17147 & 39.12 & 6.871 & 4.833 & 0.571 & 4.918 & 0.085 & 120 & 84988 & 35.67 & 7.223 & 4.985 & 0.280 & 5.214 & 0.229 \\
37 & 22632 & 15.00 & 9.280 & 5.160 & 1.294 & 5.077 & -0.083 & 121 & 85007 & 34.12 & 7.040 & 4.705 & 0.653 & 4.483 & -0.222 \\
38 & 23555 & 31.50 & 6.500 & 3.992 & -0.243 & 4.169 & 0.177 & 122 & 85042 & 51.22 & 6.505 & 5.052 & -0.081 & 4.841 & -0.211 \\
39 & 23688 & 8.79 & 9.779 & 4.499 & 1.146 & 4.177 & -0.322 & 123 & 86013 & 19.38 & 8.576 & 5.013 & 0.561 & 5.015 & 0.002 \\
40 & 24030 & 8.66 & 9.832 & 4.520 & 0.525 & 4.886 & 0.366 & 124 & 86321 & 8.38 & 9.888 & 4.504 & 0.761 & 4.365 & -0.139 \\
41 & 25860 & 20.07 & 8.774 & 5.287 & 0.356 & 5.105 & -0.182 & 125 & 87062 & 9.59 & 10.442 & 5.351 & 1.449 & 5.231 & -0.120 \\
42 & 26617 & 8.79 & 10.337 & 5.057 & 0.632 & 4.789 & -0.268 & 126 & 88945 & 40.29 & 7.098 & 5.124 & 0.265 & 4.968 & -0.156 \\
43 & 27072 & 112.02 & 3.760 & 4.006 & 0.147 & 3.889 & -0.117 & 127 & 92270 & 34.78 & 6.280 & 3.987 & 0.020 & 3.906 & -0.081 \\
44 & 28403 & 19.64 & 8.838 & 5.304 & 0.420 & 5.223 & -0.081 & 128 & 92532 & 32.72 & 7.347 & 4.921 & 0.631 & 4.705 & -0.216 \\
45 & 28671 & 16.81 & 9.511 & 5.639 & 0.871 & 5.550 & -0.089 & 129 & 92781 & 14.59 & 9.225 & 5.045 & 0.566 & 5.101 & 0.056 \\
46 & 29432 & 42.55 & 7.092 & 5.236 & 0.317 & 4.918 & -0.318 & 130 & 93185 & 41.94 & 7.009 & 5.122 & 0.545 & 4.926 & -0.196 \\
47 & 31188 & 16.79 & 8.716 & 4.841 & 0.634 & 4.846 & 0.005 & 131 & 94129 & 17.60 & 8.432 & 4.660 & -0.206 & 4.805 & 0.145 \\
48 & 33582 & 12.56 & 9.217 & 4.712 & 0.180 & 4.953 & 0.241 & 132 & 94645 & 36.30 & 6.439 & 4.239 & -0.030 & 4.074 & -0.165 \\
49 & 34017 & 52.27 & 6.163 & 4.754 & 0.087 & 4.761 & 0.007 & 133 & 96124 & 36.72 & 7.420 & 5.245 & 0.084 & 5.275 & 0.030 \\
50 & 34511 & 21.64 & 8.240 & 4.916 & 0.069 & 4.796 & -0.120 & 134 & 96258 & 39.82 & 5.884 & 3.885 & 0.149 & 3.745 & -0.140 \\
51 & 35139 & 30.95 & 8.002 & 5.455 & 0.669 & 5.370 & -0.085 & 135 & 98355 & 19.11 & 7.610 & 4.016 & 0.251 & 4.243 & 0.227 \\
52 & 36491 & 20.20 & 8.677 & 5.204 & 1.081 & 4.887 & -0.317 & 136 & 99139 & 17.95 & 9.078 & 5.348 & 0.399 & 5.113 & -0.235 \\
53 & 36849 & 12.37 & 9.109 & 4.571 & 0.547 & 4.725 & 0.154 & 137 & 99938 & 17.64 & 8.588 & 4.820 & 0.308 & 4.958 & 0.138 \\
54 & 37419 & 19.16 & 8.915 & 5.327 & 0.468 & 5.083 & -0.244 & 138 & 100017 & 56.92 & 6.083 & 4.859 & 0.341 & 4.578 & -0.281 \\
55 & 38541 & 34.30 & 8.548 & 6.224 & 1.371 & 6.236 & 0.012 & 139 & 100279 & 10.46 & 10.278 & 5.376 & 0.741 & 5.388 & 0.012 \\
56 & 38769 & 11.56 & 8.969 & 4.284 & 0.239 & 4.497 & 0.213 & 140 & 100568 & 22.78 & 8.840 & 5.628 & 1.297 & 5.310 & -0.318 \\
57 & 40118 & 45.90 & 7.116 & 5.425 & 0.188 & 5.591 & 0.166 & 141 & 100792 & 17.00 & 8.478 & 4.630 & 0.836 & 4.872 & 0.242 \\
58 & 40778 & 10.35 & 9.864 & 4.939 & 1.182 & 5.163 & 0.224 & 142 & 102018 & 24.89 & 7.433 & 4.413 & -0.298 & 4.430 & 0.017 \\
59 & 41484 & 44.94 & 6.577 & 4.840 & -0.068 & 4.902 & 0.062 & 143 & 102046 & 16.15 & 8.381 & 4.422 & 0.513 & 4.709 & 0.287 \\
60 & 42356 & 22.91 & 7.827 & 4.627 & -0.298 & 4.874 & 0.247 & 144 & 102762 & 17.11 & 8.304 & 4.470 & -0.107 & 4.566 & 0.096 \\
61 & 43595 & 7.87 & 11.031 & 5.511 & 0.850 & 5.317 & -0.194 & 145 & 102805 & 32.66 & 6.121 & 3.691 & 0.352 & 3.713 & 0.022 \\
62 & 43726 & 57.52 & 6.299 & 5.098 & -0.074 & 4.987 & -0.111 & 146 & 103458 & 45.17 & 6.746 & 5.020 & 0.391 & 5.241 & 0.221 \\
63 & 44116 & 12.70 & 8.598 & 4.117 & 0.635 & 3.981 & -0.136 & 147 & 103498 & 18.95 & 8.466 & 4.854 & 0.731 & 4.977 & 0.123 \\
64 & 44713 & 26.83 & 7.544 & 4.687 & -0.395 & 4.887 & 0.200 & 148 & 103897 & 7.70 & 10.318 & 4.750 & 0.166 & 4.893 & 0.143 \\
65 & 44811 & 24.53 & 7.829 & 4.777 & 0.515 & 4.750 & -0.027 & 149 & 104659 & 29.10 & 7.530 & 4.849 & 0.818 & 5.049 & 0.200 \\
66 & 47048 & 10.26 & 8.551 & 3.607 & 0.632 & 3.349 & -0.258 & 150 & 107294 & 9.03 & 10.155 & 4.933 & 1.219 & 4.562 & -0.371 \\
67 & 47174 & 11.99 & 10.205 & 5.599 & 0.691 & 5.380 & -0.219 & 151 & 107877 & 24.91 & 7.021 & 4.003 & 0.166 & 4.085 & 0.082 \\
68 & 48209 & 13.10 & 9.879 & 5.465 & 0.434 & 5.360 & -0.105 & 152 & 108468 & 29.93 & 7.450 & 4.831 & -0.041 & 4.876 & 0.045 \\
69 & 49285 & 21.13 & 8.367 & 4.991 & -0.197 & 5.132 & 0.141 & 153 & 108736 & 27.95 & 7.316 & 4.548 & 0.043 & 4.679 & 0.131 \\
70 & 49615 & 21.25 & 7.892 & 4.529 & 0.484 & 4.450 & -0.079 & 154 & 109144 & 19.52 & 7.412 & 3.864 & -0.266 & 4.150 & 0.286 \\
71 & 49793 & 22.58 & 8.296 & 5.065 & 0.404 & 5.186 & 0.121 & 155 & 109381 & 23.53 & 8.115 & 4.973 & -0.149 & 4.884 & -0.089 \\
72 & 49942 & 16.86 & 8.663 & 4.797 & -0.122 & 4.969 & 0.172 & 156 & 109646 & 27.64 & 7.600 & 4.808 & 0.713 & 4.647 & -0.161 \\
73 & 50834 & 8.09 & 9.369 & 3.909 & 0.585 & 3.663 & -0.246 & 157 & 110035 & 32.22 & 7.254 & 4.795 & 0.115 & 4.735 & -0.060 \\
74 & 50965 & 9.14 & 9.974 & 4.779 & 0.347 & 4.726 & -0.053 & 158 & 110341 & 32.33 & 6.294 & 3.842 & 0.302 & 3.759 & -0.083 \\
75 & 52673 & 14.07 & 9.799 & 5.540 & 0.609 & 5.451 & -0.089 & 159 & 110560 & 5.32 & 10.756 & 4.386 & 0.062 & 4.572 & 0.186 \\
76 & 52771 & 10.45 & 10.381 & 5.477 & 1.553 & 5.451 & -0.026 & 160 & 111565 & 31.82 & 7.861 & 5.375 & 0.226 & 5.523 & 0.148 \\
77 & 53537 & 20.32 & 8.188 & 4.728 & -0.131 & 4.813 & 0.085 & 161 & 112811 & 16.56 & 9.574 & 5.669 & 0.508 & 5.707 & 0.038 \\
78 & 53721 & 71.11 & 5.280 & 4.540 & -0.246 & 4.643 & 0.103 & 162 & 113688 & 12.07 & 8.837 & 4.246 & -0.408 & 4.669 & 0.423 \\
79 & 54469 & 9.27 & 9.962 & 4.797 & 0.459 & 4.707 & -0.090 & 163 & 113896 & 34.03 & 6.891 & 4.550 & -0.014 & 4.643 & 0.093 \\
80 & 54641 & 18.36 & 8.242 & 4.561 & 0.876 & 4.690 & 0.129 & 164 & 114450 & 14.48 & 8.646 & 4.450 & 0.038 & 4.516 & 0.066 \\
81 & 54924 & 15.02 & 9.248 & 5.131 & 0.606 & 4.859 & -0.272 & 165 & 114702 & 25.60 & 7.745 & 4.786 & 0.421 & 4.801 & 0.015 \\
82 & 55592 & 8.35 & 10.109 & 4.717 & 0.938 & 4.603 & -0.114 & 166 & 114924 & 48.77 & 5.762 & 4.203 & 0.038 & 4.234 & 0.031 \\
83 & 55761 & 22.95 & 8.056 & 4.860 & 0.570 & 4.680 & -0.180 & 167 & 116106 & 38.29 & 6.679 & 4.594 & 0.422 & 4.445 & -0.149 \\
84 & 56664 & 9.45 & 8.766 & 3.643 & 0.319 & 3.786 & 0.143 & 168 & 118115 & 20.84 & 8.102 & 4.696 & -0.259 & 4.889 & 0.193 \\
\hline
\end{tabular}
\end{center}
}
\end{table*}

\begin{table*}
\setlength{\tabcolsep}{4pt} 
{\scriptsize
\caption{Vertical metallicity gradients appeared in the literature for the dwarf stars. $Z$ in kpc.} 
\begin{center}
\begin{tabular}{lllll}
\hline
Authors                  & d[Fe/H]$/dZ$     & Region & Remark & Survey\\
                         & (dex kpc$^{-1}$) &        &        &       \\
\hline
\citet{Mikolaitis14}  & $-0.107\pm0.009$ & Solar neighbourhood & $0<Z<4.50$ & Gaia-ESO\\
                         & $-0.072\pm0.006$ & Solar neighbourhood &  $0<Z<4.25$ & Gaia-ESO\\
\citet{Schlesinger14} & $-0.257$         & $b>10^{\circ}$      &  $0.27\leq Z\leq1.62$ & SDSS-SEGUE\\
\citet{Peng12}        & $-0.160\pm0.060$ & $55^{\circ}\leq l \leq 260^{\circ}$, $35^{\circ}\leq |b| \leq 70^{\circ}$ &   $2<Z<5$ & BATC \& SDSS\\
\citet{Kordopatis11}  & $-0.140\pm0.050$ & $l\sim 277^{\circ}$, $b\sim 47^{\circ}$ &   $1\leq Z\leq4$ & ESO\\
\citet{Yaz10}         & $-0.320\pm0.010$ & $0^{\circ}\leq l\leq 100^{\circ}$, $140^{\circ}\leq l\leq 160^{\circ}$, $\langle b\rangle=45^{\circ}$ & $Z<2.5$ & SDSS\\
\citet{Ak07a}          & $-0.380\pm0.060$ & $0^{\circ}\leq l\leq 360^{\circ}$, $60^{\circ}\leq b\leq 65^{\circ}$ &   $3\leq Z<5$ & SDSS\\
\citet{Ak07b}          & $-0.160\pm0.020$ & $(l,b)=(180^{\circ}, 45^{\circ})$ & $Z<3$  & SDSS\\
\citet{Karaali03b}     & $-0.20$          & $(l,b)=(52^{\circ}, -39^{\circ})$ &  $Z\leq8$ & Deep Multicolor\\
\citet{Barta03}        & $-0.230\pm0.040$ & $0^{\circ}\leq l\leq 100^{\circ}$, $59^{\circ}\leq b\leq 76^{\circ}$ &  $Z<1.3$ & Vilnius\\
\hline
\end{tabular}
\end{center}
}
\end{table*}

\section{Application of the Procedure}
We estimated vertical metallicity gradients by applying our metallicity calibration to a sample of stars in a high Galactic latitude field with a series of constraints as explained in the following. Metallicity gradients can be estimated in the vertical or radial directions for the objects in our Galaxy. It has been shown that there are large numerical differences between two kind of metallicity gradients. Also, any one of these metallicity gradients (vertical or radial) is depends on many parameters such as population type, the Galactic coordinates and distance range of the objects for which the metallicity gradient is carried out \citep[cf.][]{Onal16}. Here, our data consist of dwarfs. Hence we tabulated the vertical metallicity gradients for the dwarfs appeared in the literature in Table 4 for comparison purpose. Thus, we applied the following constraints to a star sample and estimated the vertical metallicity gradients: $i$) Galactic coordinates and size: $85^{\circ}\leq b\leq 90^{\circ}$, $0^{\circ}\leq l\leq 360^{\circ}$; $A$=78 deg$^{2}$, $ii$) Luminosity class: F-G dwarfs, and $iii$) Absolute magnitude $4<M_g\leq6$ mag.   

\subsection{The star sample}
The data for the application of the procedure are provided from the recent survey DR 12 of SDSS III \citep{Alam15}. There are 1,973,575 objects with de-reddened $ugriz$ magnitudes in the star field defined above. The SDSS magnitudes and their errors are supplied from SQL webpage of SDSS\footnote{https://skyserver.sdss.org/dr12/en/tools/search/sql.aspx}. Following \citet{Chen01}, we restricted our data with $g_o\leq23$ and $(u-g)_0>0.5$ mag to avoid from the extra galactic objects. Then our sample reduced to 433,636. One sees in Fig. 10 a large scattering of these stars in the $(g-r)_0\times (r-i)_0$ two-colour diagram. We adopted the following equation of \citet{Juric08} and limited the number of stars as 411,512 which are within $\pm 2\sigma$ of this equation:

\begin{eqnarray}
(g-r)_0 = 1.39 (1-\exp[-4.9(r-i)_0^3-2.45(r-i)_0^2 \\ \nonumber
 -1.68(r-i)_0-0.050]).  
\end{eqnarray} 

\begin{figure*}[t]
\begin{center}
\includegraphics[trim=1cm 0cm 3cm 1cm, clip=true, scale=0.38]{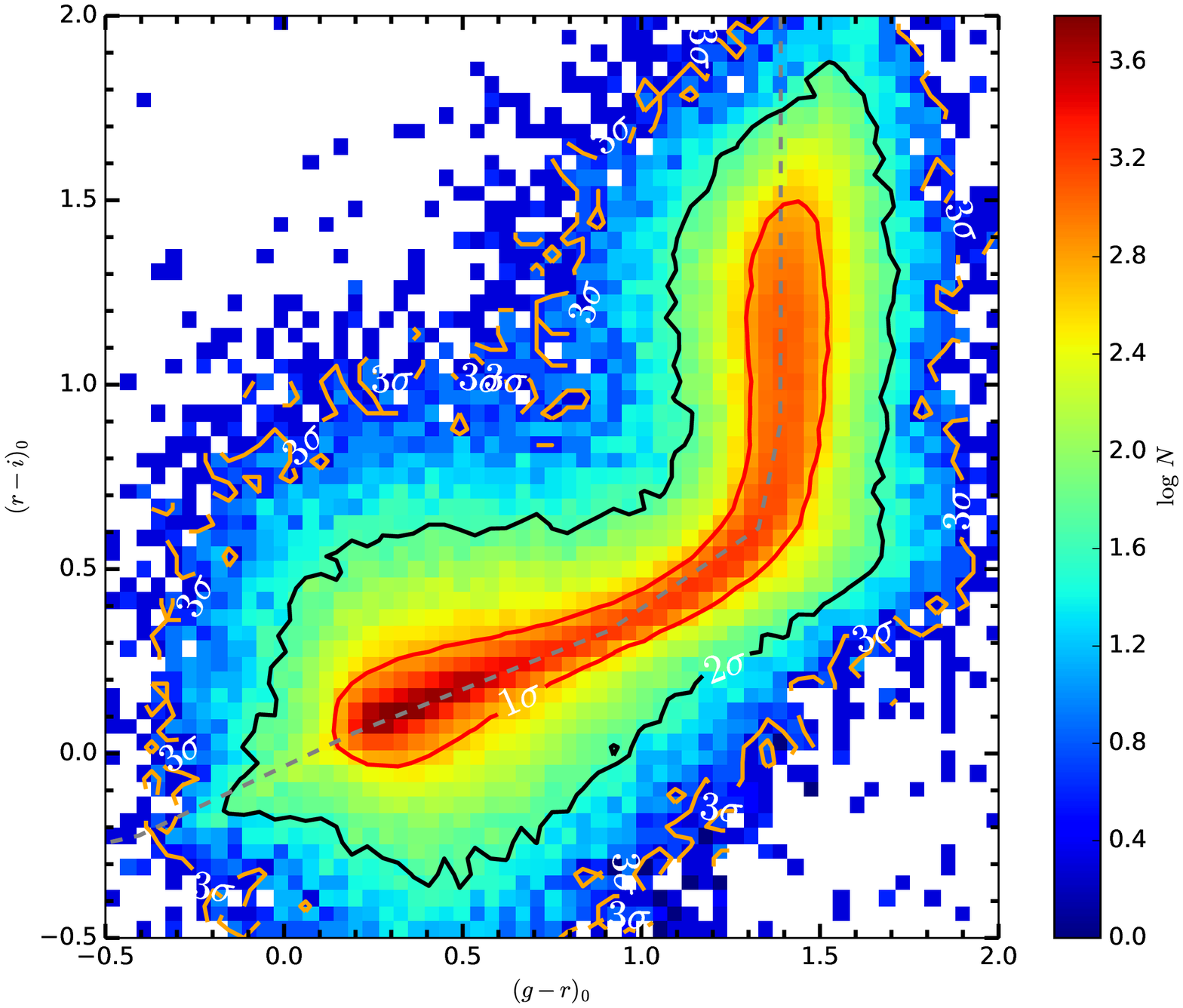}
\caption[] {$(g-r)_0 \times (r-i)_0$ two-colour diagram for 433,636 stars. The dotted mean line corresponds to the equation of \citet{Juric08}, while the lines with red, black and orange colours cover the stars within 1$\sigma$, 2$\sigma$ and 3$\sigma$ of the mean line, respectively.}
\end{center}
\end{figure*}

\begin{figure*}[t]
\begin{center}
\includegraphics[scale=0.34, angle=0]{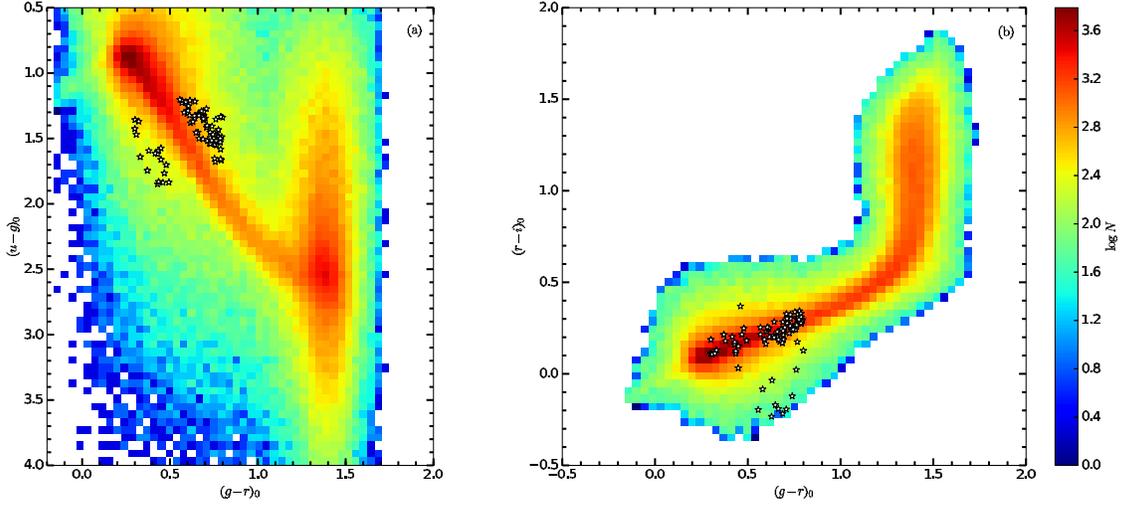}
\caption[] {$(u-g)_0 \times (g-r)_0$ and $(g-r)_0 \times (r-i)_0$ two-colour diagrams for 411,512 stars within 2$\sigma$ of the mean line. 131 giant stars are marked with black star symbol in the diagrams.}
\end{center}
\end{figure*}

\begin{figure}[h]
\begin{center}
\includegraphics[trim=1.5cm 0cm 2cm 0cm, clip=true, scale=0.38]{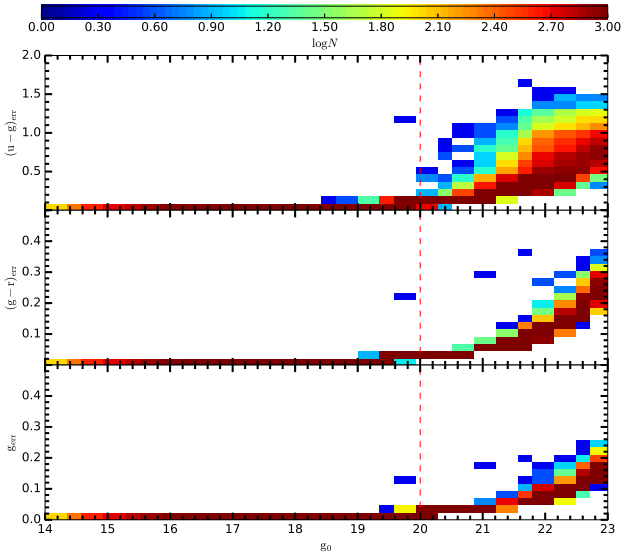}
\caption[] {Error distributions for the apparent magnitude $g$ and colours $u-g$ and $g-r$ as a function of apparent $g_0$ magnitude.}
\end{center}
\end{figure}

Finally, we used the following equations of \citet{Helmi03} \citep[see also,][]{Bilir08c} and identified 131 giants from the sample:

\begin{eqnarray}
1.1\leq(u-g)_{0}\leq2, \nonumber\\
-0.1<P_1<0.6,\\ \nonumber
|s|>m_{s}+0.05.
\end{eqnarray}
where $P_1$, $s$ and $m_s$ are defined as follows:
\begin{eqnarray}
P_{1}=0.910(u-g)_{0}+0.415(g-r)_{0}-1.28,\nonumber\\
s=-0.249u_{0}+0.794g_{0} - 0.555r_{0}+0.240, \\ \nonumber
m_s=0.12. 
\end{eqnarray}  
The position of giants are shown in the $(u-g)_0\times(g-r)_0$ and $(g-r)_0\times(r-i)_0$ two-colour diagrams of Fig. 11. Thus, our sample reduced to 411,381 dwarfs. 

We considered the errors of dwarfs in the final sample and omitted the ones with large errors from the sample to obtain more accurate results, as explained in the following. Fig. 12 shows that the errors for $g$ magnitudes and, $g-r$ and $u-g$ colours are small for a large range of $g_0$ apparent magnitude, while they increase up to 0.25, 0.35 and 1.5 mag, respectively, with increasing $g_0$. Hence, we restricted the faint limit of the de-reddened apparent magnitude with $g_0=20$ mag to avoid from large errors in magnitude and colours. Thus, the median errors for $g_0$, $(g-r)_0$, and $(u-g)_0$ for the sample of dwarfs with $14<g_0\leq 20$ ($N$=27,304) are 0.03, 0.04 and 0.11 mag, respectively.
 
Our final constraint is related to the absolute magnitudes of the stars which are estimated via the procedure in Section 3. We considered only the dwarf stars whose absolute $M_g$ magnitudes are defined in our calibrations, i.e. $4<M_g\leq6$ mag. Thus, the application of the metallicity calibration will be carried out for 23,414 dwarf stars. 

\subsection{Vertical metallicity gradient}

We estimated vertical metallicity gradients for three sub-samples with different absolute magnitudes, i.e. $4<M_g\leq6$, $4<M_g\leq5$ and $5<M_g\leq6$ mag. The number of stars in these sub-samples are 23,414, 6,833 and 16,581, respectively. The distances of the stars are evaluated by the Pogson formula and they are reduced to the vertical distances ($Z$) by the equation $Z=r\sin(b)$, where $r$ and $b$ are the distance to a star relative to the Sun and its Galactic latitude, respectively. The distance error of a star is mainly comes from its apparent and absolute magnitudes. The rms error of the absolute magnitude in our new distance calibration is 0.18 mag (see, Eq. 14). As seen in Fig. 12, the median magnitude error for the apparent magnitude $g_0=20$ becomes 0.03 mag. Thus, the errors in the distance modules of the stars in our study do not exceed 0.185 mag. This maximum magnitude error in the distance module corresponds to the distance errors of about 180, 450 and 900 pc at 2, 5 and 10 kpc.
  
The distribution of the metallicities for three sub-samples are demonstrated in Fig. 13a-c and the evaluated metallicity gradients for the distance intervals $0<Z\leq5$, $0<Z\leq2$, $2<Z\leq5$ and $Z>5$ kpc are listed in Table 5. The metallicity gradients are deep for the stars with $0<Z\leq5$, $ 0<Z\leq2$, $2<Z\leq5$ kpc in three absolute magnitude intervals, $4<M_g\leq6$, $4<M_g\leq5$ and $5<M_g\leq6$ mag. While they are all positive numbers, such as $d$[Fe/H]$/dZ = +0.046\pm0.008$ dex kpc$^{-1}$ for the stars with absolute magnitude $4<M_g\leq5$, beyond $Z=5$ kpc. The vertical metallicity gradients appeared in the literature are deepest at small $Z$ distances. Whereas, Table 5 and Fig. 13a-c show that the deepest metallicity gradient can occur at larger $Z$ distances. For instance, the deepest metallicity gradient, $d$[Fe/H]$/dZ=-0.202\pm0.029$ dex kpc$^{-1}$ in the absolute magnitude interval $4<M_g\leq6$ mag belongs to stars with small vertical distances, $0<Z\leq2$ kpc, agreeable with the results in the literature. Whereas, the deepest metallicity gradients in the absolute magnitude intervals $4<M_g\leq5$ and $5<M_g\leq6$ mag, i.e. $d$[Fe/H]$/dZ=-0.324\pm0.026$ and $d$[Fe/H]$/dZ=-0.187\pm0.013$ dex kpc$^{-1}$ respectively, are the ones for stars with larger vertical distances, $2<Z\leq5$ kpc. 
                 
\begin{figure}[h]
\begin{center}
\includegraphics[trim=2cm 2.2cm 2.5cm 0cm, clip=true, scale=0.22]{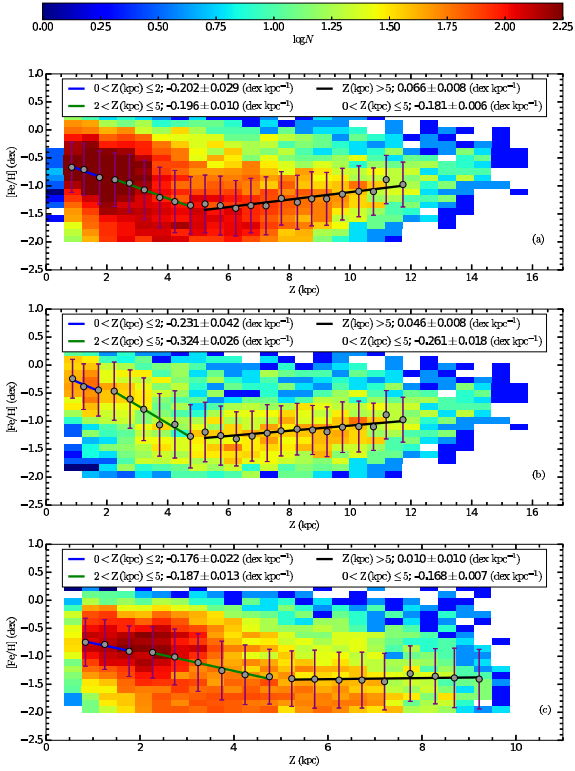}
\caption[] {Distributions of the [Fe/H] in terms of vertical distance $Z$ in three panels: for dwarf stars with (a) $4<M_g\leq6$, (b) for $4<M_g\leq5$ and (c) for $5<M_g\leq6$ mag.}
\end{center}
\end{figure}

\begin{table}
\setlength{\tabcolsep}{2pt} 
{\small
\caption{Vertical metallicity gradients, $d$[Fe/H]/$dZ$ (dex kpc$^{-1}$), as a function of vertical distance $Z$ (kpc) and absolute magnitude $M_g$ (mag) estimated in this study.} 
\begin{center}
\begin{tabular}{cccc}
\hline
$Z$  & $4<M_g\leq6$ & $4<M_g\leq5$ & $5<M_g\leq6$ \\
(kpc)  & $d$[Fe/H]$/dZ$  & $d$[Fe/H]$/dZ$ & $d$[Fe/H]$/dZ$ \\
\hline
$0<Z\leq 5$ & -0.181$\pm$0.006 & -0.261$\pm$0.018 & -0.168$\pm$0.007 \\
$0<Z\leq2$  & -0.202$\pm$0.029 & -0.231$\pm$0.042 & -0.176$\pm$0.022 \\
$2<Z\leq5$  & -0.196$\pm$0.010 & -0.324$\pm$0.026 & -0.187$\pm$0.013 \\
     $Z>5$  & +0.066$\pm$0.008 & +0.046$\pm$0.008 & +0.010$\pm$0.010 \\
\hline
\end{tabular}
\end{center}
}
\end{table}

As stated in the previous paragraph, vertical metallicity gradients appeared in the literature are deep at small vertical distances, while they are less deep or flat at large $Z$ distances. Small $Z$ distances cover the Galactic region dominated by the thin-disc population, whereas thick-disc stars are significant at larger vertical distances. Here, we showed that metallicity gradients are also absolute magnitude dependent. Actually, the vertical distance intervals of the stars with the deepest metallicity gradients in different absolute magnitude intervals are different. In the following section, we will investigate the relation between our finding and the ones already appeared in the literature.

\subsection{Identification of the metallicity gradient in term of absolute magnitude and population type}

We plotted the histogram for the stars with $4<M_g\leq6$ mag for interpretation of our results. Fig. 14 shows that the distribution is bimodal, indicating two different star categories. Actually, it could be dissolved into two different histograms with single modes by overlapping of the histograms for the stars with $4<M_g\leq5$ and $5<M_g\leq6$ mag on the same diagram. The mode for the bright absolute magnitudes correspond to $(g-r)_0\approx0.22$ mag, while for fainter ones it is at $(g-r)_0\approx0.38$ mag. We separated the sample stars into two apparent magnitude intervals, $g_0\leq18$ and $g_0>18$ mag, and repeated our procedure to obtain a detailed result. For apparently bright stars in the panel (a) of Fig. 15, the mode for the stars with absolute magnitudes $5<M_g\leq6$ almost kept its position in Fig. 14, while the other one moved from $(g-r)_0\approx0.22$ mag to $(g-r)_0\approx0.35$ mag. Contrary to the diagram in the panel (a), bimodal structure for stars with $4<M_g\leq6$ mag in the panel (b) for apparently faint stars is more evident. One can observe a slight shift of the mode for stars with $5<M_g\leq6$ mag, relative to the one in Fig. 14. While, the mode for the stars with $4<M_g\leq5$ mag is now at a position rather different than the one in panel (a), but almost equal to the one in Fig. 14, i.e. $(g-r)_0\approx0.22$ mag.
    
\begin{figure}[h]
\begin{center}
\includegraphics[scale=0.55, angle=0]{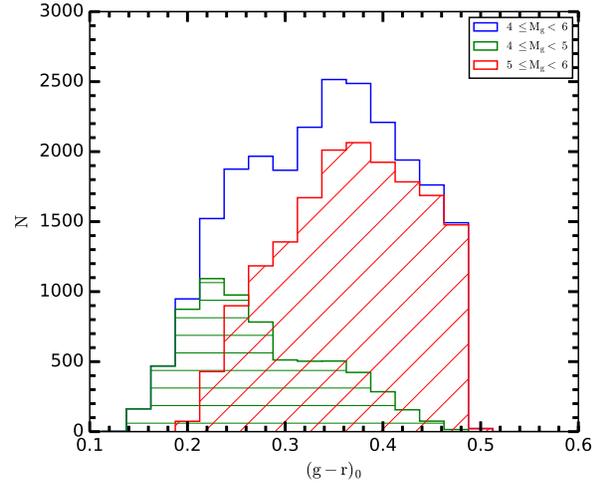}
\caption[] {Histograms for dwarf stars with absolute magnitudes $4<M_g\leq6$, $4<M_g\leq5$ and for $5<M_g\leq6$ mag.}
\end{center}
\end{figure}

\begin{figure*}[t]
\begin{center}
\includegraphics[scale=0.55, angle=0]{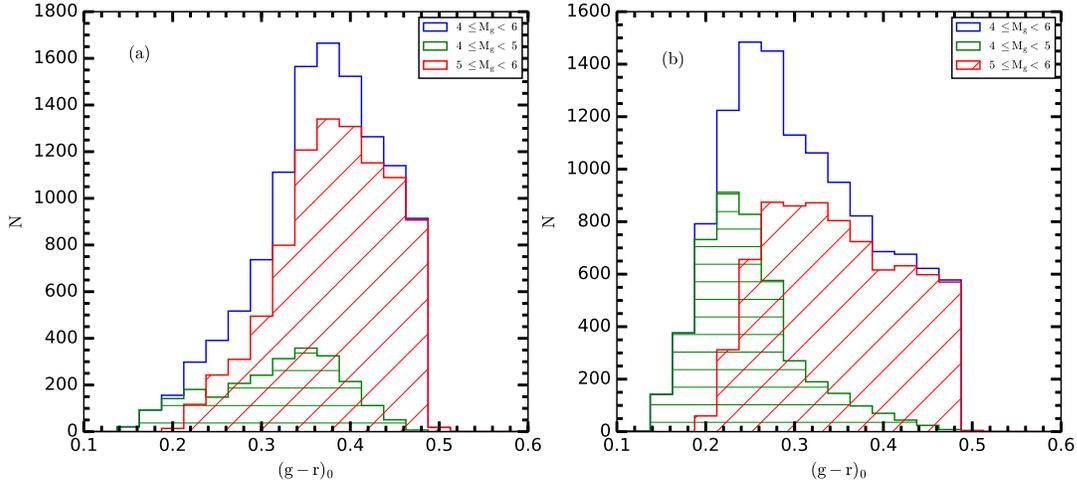}
\caption[] {The same as in Fig. 14, but for two different apparent magnitudes: (a) for $g_0\leq18$ and (b) for $g_0>18$ mag.}
\end{center}
\end{figure*}

\citet{Chen01} investigated the distribution of stars in the $g_0\times(g-r)_0$ colour magnitude diagram for two large samples north and south of the Galactic plane and revealed the positions of three different populations, i.e. thin disc, thick disc and halo. In their Fig. 6, one sees a gross bimodal structure in $(g-r)_0$, reflecting the separation of halo/thick disc at $(g-r)_0\approx0.3$ mag and the thin-disc population at $(g-r)_0\approx1.3$ mag. Our histogram in Fig. 14 is also bimodal and the mode at the blue side is a bit left of $(g-r)_0\approx0.3$ mag, while their mode at $(g-r)_0\approx1.3$ mag does not appear in our Fig. 14, due to our restrictions. However, there exist a second mode at $(g-r)_0\approx0.38$ mag. \cite{Chen01} showed also that in the range $15<g_0<18$ mag, the blue stars are dominated by thick-disc stars with a turnoff of $(g-r)_0\approx0.33$ mag, a value rather close to the colour of the mode of the histogram for the stars with $4<M_g\leq5$ mag in Fig. 15a , i.e. $(g-r)_0\approx0.35$ mag. Then, one can say that stars with apparent magnitude $g_0<18$ and absolute magnitude $4<M_g\leq5$ are dominated by thick-disc stars. In the Fig. 6 of \citet{Chen01}, for $g_0>18$ mag the Galactic halo, which has a turnoff colour $(g-r)_0\approx0.2$ mag, becomes significant. This colour is rather close to the one of the mode of the histogram for stars with $g_0>18$ mag and absolute magnitudes $4<M_g\leq5$ in Fig. 15b, $(g-r)_0\approx0.22$ mag. Then, we can say that stars with apparent magnitude $g_0>18$ and absolute magnitude $4<M_g\leq5$ are dominated by halo stars. It seems that the modes at colours $(g-r)_0\approx0.38$ mag in both panels of Fig. 15 correspond to thin-disc stars. Then, we can consider the stars with any apparent magnitude but with absolute magnitudes $5<M_g\leq6$ as thin-disc stars. We should add that additional stars with absolute magnitudes fainter than $M_g=6$ mag may shift the colour of the mode of the thin-disc stars up to $(g-r)_0\approx1.3$ mag, as in \citet{Chen01}.         

In summary, significant number of stars with absolute magnitude $5<M_g\leq6$ belong to the thin disc population. Whereas, thick disc and halo stars share the absolute magnitude $4<M_g\leq5$, i.e. stars with apparent magnitude $g_0<18$ are thick disc stars, while apparently faint ones are dominated by the halo population. Now, if we turn back to Table 5 in our study, we see that the deep metallicity gradient $-0.324\pm0.026$ dex kpc$^{-1}$ estimated for the stars with absolute magnitudes $4<M_g\leq5$, and vertical distances $2<Z\leq5$ kpc corresponds to the thick disc population. While the flat one for stars with the same absolute magnitudes but for vertical distances $Z>5$ kpc (hence for apparently fainter stars), $d$[Fe/H]$/dZ = +0.046\pm0.008$ dex kpc$^{-1}$, is a typical value for the halo population.

Two metallicity gradients for stars with absolute magnitudes $5<M_g\leq6$, $d$[Fe/H]$/dZ = -0.176\pm0.022$ and $d$[Fe/H]$/dZ = -0.187\pm0.013$ dex kpc$^{-1}$, for the consecutive vertical distance intervals $0<Z\leq2$ and $2<Z\leq2$ kpc are close the each other. Then one can say that the thin disc stars which dominate the cited absolute magnitude intervals occupy a large region in the vertical direction of our Galaxy with an intermediate vertical metallicity gradient.

\section{Summary and Discussion}
We calibrated the [Fe/H] abundances and $\Delta M_g$ absolute magnitude offsets of a sample of 168 F-G type dwarf stars in terms of reduced UV excess, $\delta_{0.41}$, defined with the $ugr$ data in SDSS. The iron abundances used in our calibrations are the updated ones used in Paper I \citep[originally  taken from][]{Bensby14, Nissen10, Reddy06, Venn04}. The $ugr$ data of the sample stars are transformed from their $UBV$ data. The distances are evaluated by using the re-reduced trigonometric parallaxes \citep{vanLeeuwen07} and the $M_g$ absolute magnitudes are evaluated by the Pogson equation. The procedures used in the calibrations of [Fe/H] and $\Delta M_g$ are the same as the corresponding ones in Paper I, i.e. each of the [Fe/H] and $\Delta M_g$ are fitted to a third order polynomial in terms of $\delta_{0.41}$, The corresponding squared correlation coefficients and standard deviations are $R^2=0.949$, $\sigma=0.107$ mag; and $R^2=0.816$, $\sigma=0.180$ mag, respectively.
       
We applied the metallicity calibration to a set of stars in a high latitude Galactic field with size 78 deg$^2$, i.e. $85^{\circ}\leq b\leq 90^{\circ}$, $0^{\circ}\leq l\leq 360^{\circ}$. Our calibrations are restricted with absolute magnitudes $4<M_g\leq6$. Hence we considered only the stars, totally $N=23,414$, which supply our restriction in absolute magnitude. Also, we limited the apparent magnitudes of the stars (dwarfs) with $14<g_0\leq20$ mag to avoid from large magnitude and colour errors. 

We estimated metallicity gradients as a function of vertical distance, $0<Z\leq5$, $0<Z\leq2$, $2<Z\leq5$, $Z>5$ kpc, for stars with absolute magnitude, $4< M_g\leq6$, $4<M_g\leq5$ and $5<M_g\leq6$. Some of our results (Table 5, Fig. 13) confirm the ones in the literature, however some of them need explanation. Vertical metallicity gradients for dwarfs with $Z>5$ kpc in our study are all positive numbers for all absolute magnitude intervals, agreeable with the expectation that the vertical metallicity gradient is not deep at large $Z$ distances. Also, in the absolute magnitude interval $4<M_g\leq6$, the gradient $d$[Fe/H]$/dZ=-0.202\pm0.029$ dex kpc$^{-1}$ for $0<Z\leq2$ kpc is deeper than the one for larger $Z$ distances, $2<Z\leq5$, i.e. $d$[Fe/H]$/dZ=-0.196\pm0,010$ dex kpc$^{-1}$, again agreeable with the findings appeared in the literature. Whereas, the case is different for $d$[Fe/H]$/dZ=-0.231\pm0.042$ and $d$[Fe/H]$/dZ=-0.324\pm0.026$ dex kpc$^{-1}$ in the absolute magnitude interval $4<M_g\leq5$, in the sense that the deeper gradient corresponds to larger $Z$ distances, $2<Z\leq5$ kpc. This is new and it needs explanation. In Section 4, we showed that thick disc and halo stars share the absolute magnitudes $4 <M_g\leq5$. It seems that the thick-disc stars dominate the vertical distance interval $2<Z\leq5$ kpc with a gradient of $d$[Fe/H]$/dZ=-0.324\pm0.026$ dex kpc$^{-1}$, while the halo stars which are apparently fainter ($g_0>18$ mag) are significant at larger $Z$ distances with gradient $d$[Fe/H]$/dZ=+0.046\pm0.008$ dex kpc$^{-1}$ (Table 5 and Fig. 13b). As we stated in Section 4, the absolute magnitudes $5<M_g\leq 6$ are dominated with the thin-disc stars with a mode at $(g-r)_0\approx0.38$ mag. On the other hand, the deepest gradient in this absolute magnitude interval, $d$[Fe/H]$/dZ=-0.187\pm0.013$ dex kpc$^{-1}$, corresponds to the vertical distance interval $2<Z\leq5$ kpc. However it is rather close to the one in $0<z<\leq2$ kpc, $d$[Fe/H]$/dZ=-0.176\pm0.022$ dex kpc$^{-1}$. It seems that sufficient number of thin-disc stars lie up to vertical distance $Z=5$ kpc.  

Now, we will compare the vertical metallicity gradients estimated in this study with the ones in Table 4. We will prefer the star samples with high Galactic latitudes. Doing this, we will adopt the stars with the absolute magnitudes $5<M_g\leq6$ as thin-disc stars, and those with $4<M_g\leq5$ mag and $Z\leq5$ kpc as thick-disc stars, while stars with $4<M_g\leq6$ mag will be adopted as the combination of two populations. For thin disc: the gradient in \citet{Schlesinger14}, $d$[Fe/H]$/dZ=-0.243^{+0.039}_{-0.053}$ dex kpc$^{-1}$ and one in \citet{Yaz10}, $d$[Fe/H]$/dZ=-0.320\pm 0.010$ dex kpc$^{-1}$ are both deeper than ours. For thick disc: the gradient in \citet{Ak07a}, $d$[Fe/H]$/dZ=-0.380\pm0.060$ dex kpc$^{-1}$ is a bit deeper than $d$[Fe/H]$/dZ=-0.324\pm0.026$ dex kpc$^{-1}$ in our study. The gradient in \citet{Karaali03b}, $d$[Fe/H]$/dZ=-0.200$ dex kpc$^{-1}$ for thin and thick disc populations is compatible with the three gradients estimated for stars with absolute magnitudes $4<M_g\leq6$, $d$[Fe/H]$/dZ=-0.181\pm0.006$, $d$[Fe/H]$/dZ=-0.202\pm0.029$ and $d$[Fe/H]$/dZ=-0.196\pm0.010$ dex kpc$^{-1}$. 

{\bf Conclusion:} Metallicity gradients are vertical distance and absolute magnitude dependent. The thin-disc stars cover the absolute magnitude interval $5<M_g\leq6$. Whereas the thick disc and halo stars share the absolute magnitudes $4<M_g\leq5$, i.e. apparently bright ones ($g_o\leq18$ mag) are thick-disc stars, while the faint stars ($g_o>18$ mag) are dominated by halo population.    

\section{Acknowledgments}

We would like to thank the referee Dr. C. Du for useful and constructive comments concerning the manuscript.
This work has been supported in part by the Scientific and
Technological Research Council (T\"UB\.ITAK) 114F347. Part of this
work was supported by the Research Fund of the University of Istanbul,
Project Number: 52265. This research has made use of the SIMBAD, 
and NASA\rq s Astrophysics Data System Bibliographic Services.

\end{document}